\documentclass{emulateapj}
\usepackage{graphicx}
\usepackage{CJK}
\usepackage{tablefootnote}
\usepackage{subfigure}
\usepackage{multirow}

\newcommand{\xmm}{{\em XMM-Newton }}

\begin{document}
\title{A sparkler in the fireworks galaxy: discovery of an ultraluminous X-ray transient with a strong oxygen line in NGC\,6946}
\author{Chen Wang\altaffilmark{1}, Roberto Soria \altaffilmark{2,3,4}, Junfeng Wang\altaffilmark{1}}

\altaffiltext{1}{Department of Astronomy and Institute of Theoretical Physics and Astrophysics, Xiamen University, Xiamen, 361005, China}
\altaffiltext{2}{College of Astronomy and Space Sciences, University of the Chinese Academy of Sciences, Beijing 100049, China}
\altaffiltext{3}{National Astronomical Observatories, Chinese Academy of Sciences, Beijing 100012, China}
\altaffiltext{4}{Sydney Institute for Astronomy, School of Physics A28, The University of Sydney, Sydney, NSW 2006, Australia}

\email{jfwang@xmu.edu.cn}

\begin{abstract}
We discovered and studied an ultraluminous X-ray source (CXOU J203451.1$+$601043) that appeared in the spiral galaxy NGC\,6946 at some point between 2008 February and 2012 May, and has remained at luminosities $\approx$2--4 $\times 10^{39}$ erg s$^{-1}$ in all observations since then. Our spectral modelling shows that the source is generally soft, but with spectral variability from epoch to epoch. Using standard empirical categories of the ultraluminous regimes, we find that CXOU J203451.1$+$601043 was consistent with a broadened disk state in 2012, but was in a transitional state approaching the super-soft regime in 2016, with substantial down-scattering of the hard photons (similar, for example, to the ultraluminous X-ray source in NGC\,55). It has since hardened again in 2018--2019 without any significant luminosity change. The most outstanding property of CXOU J203451.1$+$601043 is a strong emission line at an energy of of $(0.66 \pm 0.01)$ keV, with equivalent width of $\approx$100 eV, and de-absorbed line luminosity of $\approx$2 $\times 10^{38}$ erg s$^{-1}$, seen when the continuum spectrum was softest. We identify the line as O {\footnotesize{VIII}} Ly$\alpha$ (rest frame energy of 0.654 keV); we interpret it as a strong indicator of a massive outflow. Our finding supports  the connection between two independent observational signatures of the wind in super-Eddington sources: a lower temperature of the Comptonized component, and the presence of emission lines in the soft X-ray band. We speculate that the donor star is oxygen-rich: a CO or O-Ne-Mg white dwarf in an ultracompact binary. If that is the case, the transient behaviour of CXOU J203451.1$+$601043 raises intriguing theoretical questions.

\end{abstract} 

\section{Introduction}

Off-nuclear, point-like sources with X-ray luminosities $>$10$^{39}$ erg s$^{-1}$  are generally interpreted as the high-luminosity tail of stellar-mass X-ray binaries \citep{kaaret17,feng11}. As such, they must be emitting at or above their Eddington limit, which is $\approx$2 $\times 10^{38}$ erg s$^{-1}$ for an accreting neutron star (NS) or $\approx$1.3 $\times 10^{39}$ erg s$^{-1}$ for a 10~M$_{\odot}$ black hole (BH); this justifies the name ``ultraluminous X-ray sources'' (ULXs) given to that sub-population. The identification of ULXs as stellar-mass X-ray binaries (most likely, with a massive donor star) is based both on studies of individual sources, in rare cases where  the compact object can be identified as a neutron star \citep{bachetti14,motch14,furst16,israel17a,israel17b,carpano18,sathyaprakash19,castillo19}, and on their statistical population properties, which are consistent with an extension of the high-mass X-ray binary luminosity function \citep{swartz11,mineo12,lehmer19}. This is also the reason why the majority of ULXs are found in star-forming galaxies rather than old ellipticals  \citep{swartz04,earnshaw19a}. It is still possible that a very small fraction of ULXs \citep{wiersema10,sutton12,earnshaw19a}, especially those located in dwarf galaxies  \citep{moran14,mezcua18}, are sub-Eddington intermediate-mass BH, but we are not concerned with that population here.





Several physical properties of ULXs either remain unexplained, or are consistent with  more than one scenario. In this paper, we focus on two unanswered questions. The first question is the properties of transient ULXs. A census of transient ULXs is largely incomplete because of the relatively short history of X-ray observations and sparse monitoring cadence of individual galaxies with ULXs. For sub-Eddington X-ray binaries \citep{remillard06}, the rule of thumb is that systems with low mass donors, accreting via Roche lobe overflow, are generally transient, while wind accreting systems with a high mass donor are generally persistent X-ray sources. For ULXs, we still cannot confidently predict which systems may exhibit a thermal-viscous instability (invoked to explain the outburst cycle in sub-Eddington sources), or what other mechanisms ({\it e.g.}, accretor/propeller transitions, or precession of the polar funnel in and out of our line of sight) may cause an apparently transient behaviour. Transient ULXs have been observed in early-type galaxies (M\,86: \citealt{vanhaaften19}; NGC\,5128: \citealt{burke13}), but more often in spiral galaxies (M\,31: \citealt{middleton12}; M\,83: \citealt{soria12}; M\,101: \citealt{kuntz05}; NGC\,5907: \citealt{walton15,pintore18}; the pulsar ULX in NGC\,300: \citealt{carpano18}) and even in starburst galaxies (NGC\,3628: \citealt{strickland01}; the pulsar ULX in M\,82: \citealt{feng07}). We do not know yet whether there are particular types of donor stars or age ranges that are more likely to be associated with transient ULXs. This is, in turn, part of the more general identification problem of the donor star in any ULX. Even in systems where a point-like UV/optical/IR counterpart is clearly identified, it is often difficult to distinguish the contributions from the donor star, the irradiated accretion disk, disk outflows, and perhaps circumbinary material \citep{tao11,gladstone13,heida14,fabrika15,lopez17,lau19}.

The second issue we discuss in this paper is the spectral evolution of individual ULXs. Individual and population studies have shown clear evidence of X-ray spectral and time-variability differences between softer and harder ULXs \citep{sutton13}. There is also solid theoretical \citep{poutanen07,ohsuga11,kawashima12,narayan17} and observational \citep{pinto16,pinto17,walton16,kosec18} evidence that super-Eddington sources launch massive, radiation-driven winds.  These winds are expected to have a lower-density funnel along the polar direction and a higher optical depth at higher viewing angles (closer to the disk plane). The wind downscatters the harder photons emitted from the innermost part of the flow, introducing a characteristic spectral curvature and high-energy downturn, and enhancing short-term X-ray variability. Based on those findings, it is now commonly accepted that softer and harder ULX spectral shapes correspond to sources seen through a thicker or thinner wind, respectively \citep{middleton15a,middleton15b}. However, it is not clear whether those differences can be explained mostly as an effect of our viewing angle (a similar scenario to the so-called Active Galactic Nuclei unification model), or instead correspond to intrinsically different physical regimes, with different wind properties. To reduce this ambiguity, more discoveries and studies of spectral state changes in individual sources would be very useful (in parallel to population studies of spectral differences between different sources). To date, spectral state transitions have been well studied (and related to wind properties) only in a handful of ULXs; for example in NGC\,247 X-1 \citep{feng16}, NGC\,55 X-1 \citep{pinto17}, IC\,342 X-1 \citep{shidatsu17}, and a few other examples discussed in \cite{pintore17} and \cite{weng18}.



Given the variety of alternative scenarios for spectral evolution in ULXs, and the limited number of X-ray state transitions identified in individual systems so far, any new identifications of such behaviour can help constrain the models. In this paper we report on our discovery of transient behaviour, spectral evolution and wind signatures in a ULX (CXOU J203451.1$+$601043) in NGC\,6946 (``the fireworks galaxy'');  four other ULXs have already been identified in this galaxy \citep{earnshaw19b}.

CXOU J203451.1$+$601043 was undetected in multiple {\em Chandra}, {\em XMM-Newton} and {\it Swift} observations between 2001 and 2008; it was first detected by {\em Chandra} in 2012, and has subsequently remained in the ultraluminous state. We monitored its behaviour using the long series of X-ray observations that have covered the highly star-forming host galaxy NGC\,6946. We searched for an optical counterpart using archival {\em {Hubble Space Telescope}} ({\it{HST}}) images.  For the host galaxy, we assumed the recently determined distance of $7.7 \pm 0.3$ Mpc \citep{anand18,eldridge19}, higher than the distances of $\approx$5.5 Mpc \citep{tully88} or $\approx$5.9 Mpc \citep{karachentsev00} commonly used in the literature before 2018; for example, the new distance implies an increase in the intrinsic source luminosites by a factor of $\approx$1.7, when we compare our results with the {\it Chandra} study of \citet{2008ApJS..177..465F}.

The paper is organized as follows. Observations and data analysis are described in Section~2. In Section 3, we present the X-ray spectral and time variability properties, and constrain a possible optical counterpart. In Section~4, we discuss how this system fits in our current knowledge of ULXs, in particular those with evidence of a strong wind, and we propose a white dwarf scenario for the donor star. Finally, prospects for follow-up studies are summarized in Section~5.
 
\section{Observations and data analysis}

\subsection{{\it Chandra}}

NGC\,6946 was observed by {\it {Chandra}} nine times between 2001 and 2017 (in most cases, to follow supernova explosions, which occur quite frequently in this galaxy).  The observation log is shown in Table~\ref{tab1}. All observations were made with the S3 chip of the Advanced CCD Imaging Spectrometer array (ACIS), except for the ACIS observation of 2012 (ObsID 13435) in which the target was placed on the S2 chip. 
After downloading the data from the public archive, we reprocessed and analyzed them with the Chandra Interactive Analysis of Observations ({\sc {ciao}}) software version 4.10 \citep{fruscione06}, with calibration database version 4.7.9. Specifically, we rebuilt level-2 event files with the task {\it {chandra\_repro}}, and filtered out background flares with the task {\it{deflare}}. We created images in multiple energy bands for each epoch with {\it {dmcopy}}. We searched for point sources in each epoch with with {\it {wavdetect}}.  Our target was undetected at every epoch until its first appearance in 2012 May. It was later detected also in the subsequent {\it Chandra} observations of 2016 September and 2017 June.

In order to check whether the source is spatially extended, we computed the point spread functions (PSFs) at the off-axis locations of the source in  2012 and 2016, using {\em Chandra} Ray Tracing (ChaRT\footnote{\url{http://cxc.harvard.edu/ciao/PSFs/chart2/index.html}}), and simulated PSF files with the {\sc marx} software\footnote{\url{http://space.mit.edu/CXC/MARX}}.  
We compared the profiles of the source with those of the respective simulated PSFs with the {\sc ciao} tool {\it srcextent}. We confirm that the transient source is consistent with being point-like. This is further confirmed by the 2017 observation (ObsID 19040), the only one in which the target is almost on-axis: its observed full-width half-maximum of $\approx$2$^{\prime\prime}$ is consistent with the expected with of an on-axis PSF.

For each observation (both those with a detection and those with no detection), we extracted the source events from a circular aperture of either $5^{\prime\prime}$ radius or $2^{\prime\prime}$ (for ObsID 19040), to match the size of the PSF at the respective locations. Background events were extracted from nearby source-free regions at least three times the size of the source region. We used the {\sc ciao} task {\it scrflux} to estimate an upper limit to the source flux in the epochs when it was not detected, and an approximate flux in the two epochs (2012 and 2017) when it was detected but with a small number of counts. For the 2016 dataset, we combined the spectra from the two exposures taken on 2016 September 28, and obtained enough counts for meaningful spectral fitting; we extracted source and background spectra with {\it specextract}, which also generates the appropriate auxiliary response files (ARFs) and response matrix files (RMFs) for the subsequent spectral analysis. Spectra were grouped to a minimum of 15 counts per bin, for $\chi^2$ fitting.  {We also repeated our spectral analysis on the same spectra grouped to 1 count per bin, using the Cash statistics \citep{cash79}.}
Moreover, to check for short term variability in each observation in which the source was detected, we extracted background subtracted light curves using the {\sc ciao} tool {\it dmextract}. 

We did subsequent data analysis with NASA's High Energy Astrophysics Software (HEASOFT): {\sc ds9} \citep{joye03} version 8.0 for imaging and photometry, {\sc ftools}/{\it Xronos} \citep{blackburn95} version 6.25 for timing analysis, and {\sc xspec} \citep{arnaud96} version 12.9.1 for spectral modelling. The reported errors are 90\% confidence intervals for the fitting parameters.  For the 2016 {\it Chandra} spectra, all the best-fitting parameters obtained from $\chi^2$ fitting and from Cash-statistics fitting ({\it cstat} in {\sc xspec}) agree well within their error ranges; thus, in Section 3.3 and Table 2 we report only the values from $\chi^2$ fitting, for simplicity.

\subsection{{\it XMM-Newton}}

There are fourteen \xmm observations of NGC\,6946 with the European Photon Imaging Camera (EPIC) in the full-frame mode. We downloaded the data from the {\it XMM-Newton} Science Archive archive and reduced them with the Science Analysis Software ({\sc sas}) version 17.0. The EPIC-pn and EPIC-MOS events were processed using {\it XMM-Newton} pipeline and the associated calibration files. We removed time intervals of high background from the event files with {\it evselect}, using a ``RATE$<=$0.4'' threshold for the pn and ``RATE$<=$0.35'' for MOS1 and MOS2. 

CXOU J203451.1$+$601043 was not detected in any of the observations between 2003 and 2007.  We grouped the non-detection observations by year to increase the signal to noise ratio. For each year, we built a stacked pn and MOS image and measured the total counts inside a circle with $20^{\prime\prime}$ radius at the position of the source, and the background counts from surrounding regions. We then applied the Bayesian method of \cite{kraft91} for Poisson-distributed counts to obtain the 90\% upper limit to the net counts and count rates\footnote{ The confidence interval tables and plots provided by \cite{kraft91} cover only the range of $\approx$0 to 10 counts. The (raw) source and background counts in our EPIC images are typically higher than that. To obtain the Bayesian confidence intervals for our data, we used the Bayesian Analysis Toolkit package \citep{caldwell09} version 1.0.0, downloaded from https://bat.mpp.mpg.de.}. For the conversion from count rates to fluxes, we used the EPIC exposure maps to account for vignetting, and we assumed that for an equal effective exposure time between the three instruments, the pn contributes to about 61\% of the count rate and MOS1+MOS2 to about 39\%. We obtained this relative ratio between the EPIC instruments with the Portable, Interactive Multi-Mission Simulator ({\sc pimms}) version 4.9, for a range of plausible spectral models; the relative contribution of pn and MOS changes at most by 2 or 3 per cent from model to model, which is negligible for the purpose of our analysis. We converted the upper limits on the pn$+$MOS count rates to flux upper limits with PIMMS, using a power-law model with photon index $\Gamma = 2.5$ and total absorbing column density $N_{\rm H} = 4 \times 10^{21}$ cm$^{-2}$ (Table 1).

In contrast, CXOU J203451.1$+$601043 was detected in 2012 and 2017, with luminosities exceeding $10^{39}$ ergs s$^{-1}$. For the 2012 and 2017 detections, source events were extracted from a circular aperture of $20^{\prime\prime}$ radius. We generated background-subtracted lightcurves  binned to 0.1 s, with the {\sc sas} task {\it evselect} followed by {\it epiclccorr}. Background events were extracted from point source-free, circular regions on the same chip approximately three times larger than the source region. We created spectral files and associated instrumental responses with the script {\it multiespecget}, which extracts the spectra of all three detectors and combines them into a single EPIC spectrum. We chose to combine the three EPIC spectra to increase the signal-to-noise ratio of possible line features in the soft X-ray band. For spectral extraction, we used the standard filtering conditions ``(FLAG==0) \&\& (PATTERN$<=$4)'' for the pn, and ``(\#XMMEA\_EM \&\& (PATTERN$<=$12)'' for MOSs.  Spectra were grouped to a minimum of 25 counts per bin, for $\chi^2$ fitting.

As for the {\it Chandra} data, subsequent imaging, timing and spectral analysis was carried out with HEASOFT packages ({\sc ds9}, {\sc ftools}, and {\sc xspec}), and with {\sc astropy} \citep{astropy18} specifically for period searches (Section 3.2).

\subsection{{\it Swift} and {\it NuSTAR}}

The X-ray Telescope (XRT) onboard {\it Swift} has monitored NGC\,6946 frequently over the years, but typical exposure times are very short ($\sim$1 ks). For this work, we used all the X-Ray Telescope observations from 2008 onwards, stacked into datasets for individual years (Table 1). The 2008 observations are the last ones from any X-ray observatory in which the transient ULX is not detected; it is detected in all subsequent {\it Swift} observations. We used standard HEASOFT packages (version 6.25) for spectral extraction: we created a combined spectrum and exposure map with {\it xselect}, and an ancillary response function with {\it xrtmkarf}; the ready-made response file comes from the XRT Calibration Database\footnote{https://swift.gsfc.nasa.gov/proposals/swift\_responses.html.}. 

We also searched for {\em NuSTAR} observations that covered the position of the transient, and found two: one from 2017 May (obsIDs 90302004002, 66.8 ks) and the other from 2017 June (obsIDs 90302004004, 47.8 ks). We examined the Focal Plane Module A and Focal Plane Module B data, processed by the pipeline task {\it nupipeline}; however, we did not detect any source at the ULX position,  neither in the individual images, nor in a stack of the two images.

\subsection{{\it HST}}

The field containing our target source was observed (Table 4) with the {\em Advanced Camera for Surveys} (ACS) Wide Field Channel (WFC) on 2004 July 29, with the F814W filter (exposure time of 120 s). It was then re-observed with ACS-WFC on 2016 October 26, in the F606W and F814W filters (exposure times of 2430 s and 2570 s, respectively). It was later imaged with the Wide Field Camera 3 (WFC3), Ultraviolet and VISible light camera (UVIS), on 2018 January 5, in the F555W band, for 710 s, and in F814W, for 780s.  

We retrieved calibrated, geometrically-corrected images (.drc files) from the Mikulski Archive for Space Telescopes. We used {\sc ds9} to inspect the images and perform aperture photometry of the candidate counterparts, with a source extraction radius of 0$''$.15 and a local annular background region. We then converted these small-aperture measurements to infinite-aperture values with the help of the online tables of encircled energy fractions for ACS-WFC and WFC3-UVIS. Finally, we applied the corresponding zeropoints, to convert our photometric measurements into magnitudes in the Vega system.

\section{Results}

\subsection{Detection and location of the X-ray transient}

The first detection of this transient as a bright new X-ray source was in the {\it Chandra} observations of 2012 May; its luminosity was two orders of magnitude higher than typical previous non-detection limits (Figure 1, and Table 1). It has remained very luminous in all subsequent {\it Chandra}, {\it XMM-Newton} and {\it Swift} observations, including the most recent ones (Figure 2). The transient resides in one of the spiral arms (Figure~\ref{optical}); the projected galactocentric radius is $\approx$90$''$, which is $\approx$3.4 kpc at the assumed distance of 7.7 Mpc. 

At first, we improved the astrometry of the stacked {\it Chandra}/ACIS images with the help of a small sample of sources that have a counterpart in the {\it Gaia} and 2MASS catalogs (available in {\sc ds9}). We estimate that our transient source is located at R.A.(J2000) $= 20^h 34^m 51^s.1$, Dec.(J2000) $=60^{\circ}10^{\prime}43^{\prime\prime}.6$, but the uncertainty remained large, $\approx$0$^{\prime\prime}.5$. The reason for this large positional error is caused by the off-axis location of the X-ray source (hence, a distorted PSF) in three of the four observations in which it is detected; the only {\it Chandra} observation in which the ULX is almost on-axis is also short (ObsID 19040, 10 ks) and there are no direct X-ray/Gaia or X-ray/2MASS associations for that short exposure alone. 

Therefore, we used a different method in our search for an optical counterpart. In the on-axis observation 19040, there are two bright X-ray sources within 2$^\prime$ of the transient ULX, with a well-identified, point-like optical counterpart in the {\it HST} images. One is SN\,2017eaw \citep{wiggins17}, located at R.A.(J2000) $= 20^h 34^m 44^s.24$, Dec.(J2000) $=60^{\circ}11^{\prime}35^{\prime\prime}.9$; the other is the ULX inside the MF16 nebula \citep{roberts03}\footnote{The MF16 nebula is obviously extended, but the optical counterpart for the peak of the X-ray emission is a point-like, blue star in the centre of the nebula.}. We determined the centroids of the X-ray emission from the three sources, with {\it wavdetect} applied to the ACIS image from ObsID 19040. The uncertainty in the centroid position is $\lesssim$0$^{\prime\prime}.1$ for each of the three sources. The transient ULX is located 51$^{\prime\prime}$.4 east and 52$^{\prime\prime}$.6 south of SN\,2017eaw. It is also located 71$^{\prime\prime}$.8 west and 47$^{\prime\prime}$.3 south of the MF16 ULX. 

Using those relative offsets in the {\it HST} images, we constrained the error circle of the transient ULX. SN\,2017eaw (but not MF16) is in the field of view of the WFC3 image from 2018; its relative offset gives the transient ULX location in those images. In the 2004 and 2016 ACS images, both MF16 and the position of SN\,2017eaw are in the field of view\footnote{SN\,2017eaw was obviously not visible at those epochs, but its precise location on the ACS chip is easily determined from the relative position of the surrounding stars, compared with the 2018 images.}; this gives us two reference offsets for the relative position of the transient ULX. Both offsets point to the same location with a difference of $<$0$^{\prime\prime}$.1 between them. In summary, we constrained the relative position of the transient ULX on the {\it HST} images with an error radius of $\lesssim$0$^{\prime\prime}$.2 

Finally, we also refined the absolute astrometry of the {\it HST} images, based on {\it Gaia} and Sloan Digital Sky Survey associations. This was a much simpler and straight-forward task, and reduced the uncertainty on the absolute astrometry of the {\it HST} images to $\lesssim$0$^{\prime\prime}.1$.  
The most accurate position for the transient ULX is then 
R.A.(J2000) $= 20^h 34^m 51^s.12$, Dec.(J2000) $=60^{\circ}10^{\prime}43^{\prime\prime}.3$ ($\pm 0^{\prime\prime}$.2).

\subsection{X-ray lightcurve}

First, we studied the long-term X-ray variability of CXOU J203451.1$+$601043. We determined the net count rate or 90\% upper limit in each {\it Chandra}/ACIS observation,  in the 0.3--7 keV band. In some cases, we stacked observations taken a few weeks apart, to reach a deeper detection limit. We converted  0.3--7 keV count rates to 0.3--10 keV fluxes assuming the same model for all {\it Chandra} observations: a power-law with photon index $\Gamma = 2.5$ and intrinsic neutral-absorption column density $N_{\rm H} = 2 \times 10^{21}$ cm$^{-2}$, in addition to the line-of-sight Galactic $N_{\rm H} = 2 \times 10^{21}$ cm$^{-2}$. We chose these model parameters because they are a good approximation to those found from a detailed fit to the 2016 {\it Chandra} data (Table 2 and Section 3.3). If we use a ``standard'' photon index $\Gamma = 1.7$ and only line-of-sight absorption, the inferred luminosities or upper limits will be $\approx$75\% of those reported in Table 1.  

We did a similar analysis for the {\it XMM-Newton}/EPIC observations, stacking the exposures from individual years to increase the signal-to-noise ratio.  Count rates were extracted from the 0.3--10 keV band.  Count rates were extracted from the 0.3--10 keV band. For the 2012 and 2017 observations, we had enough counts to fit multi-component models to the data (as we shall discuss in Section 3.3). We adopted the best-fitting Comptonization model to convert from net count rates to the fluxes and luminosities listed in Table 1. For the {\it XMM-Newton} observations in which the source was not detected, we determined upper count-rate limits from the combined EPIC images, and we used our fiducial power-law model ($\Gamma = 2.5$ and total column density $N_{\rm H} = 4 \times 10^{21}$ cm$^{-2}$) to convert to flux and luminosity limits. 

For the stacked {\it Swift}/XRT observations from 2008 (total of $\approx$10 ks between February 4 and February 14) and 2013 (total of $\approx$7 ks between May 31 and June 4), we used our fiducial power-law model ($\Gamma = 2.5$, $N_{\rm H} = 4 \times 10^{21}$ cm$^{-2}$) to determine fluxes and luminosities, or their upper limits. For the other three sets of stacked {\it Swift} observations (44 ks in 2017, 43 ks in 2018, and 17 ks so far in 2019),  we had enough counts to determine the hardness ratios between the 1.5--10 keV band and the 0.3--1.5 keV band. We fixed the total column density to $N_{\rm H} = 4 \times 10^{21}$ cm$^{-2}$, and used {\sc pimms} to estimate the power-law photon indices that most closely reproduce the observed hardness ratios; the values are $\Gamma = 3.2 \pm 0.5$ in 2017, $\Gamma = 2.1 \pm 0.4$ in 2018, and $\Gamma = 2.6 \pm 0.5$ in 2019. We then used those photon indices to convert from net count rates to fluxes and luminosities in the respective observations.

The resulting long-term lightcurve is shown in Figure~\ref{long-lc}. CXOU J203451.1$+$601043 was always undetected in all observations between 2001 and 2008, and always detected from 2012 to 2019. The upper limit to the non-detections are typically $\approx$few $\times 10^{37}$ erg s$^{-1}$ for individual years, and $<$10$^{37}$ erg s$^{-1}$ if all the observations with non-detections are stacked up. We also know that the source was undetected in {\it ROSAT} High Resultion Imager in 1994 May\citep{schlegel00}, with an upper limit to the de-absorbed 0.3--10 keV luminosity of $\approx$10$^{38}$ erg s$^{-1}$ (after converting from the model and distance used in that paper to those used for this work). Since 2012, the source has been hovering at luminosities $\approx$1--4 $\times 10^{39}$ erg s$^{-1}$ (depending on the choice of spectral model).   

We also determined and examined the background-subtracted lightcurves from individual {\em Chandra} and \xmm observations with detections; for this, we used the {\sc ftools} tasks {\it lcurve} and {\it lcstats}. We found that the 2016 {\em Chandra} observations, binned to 500 s, show statistically significant intra-observation variability by a factor of 2 (Figure~\ref{short-lc}, top panels), with a $\chi^2$ probability of constant rate $<$1\%. The intra-observational variability during the 2012 and 2017 {\it XMM-Newton} observations is more marginal (Figure~\ref{short-lc}, bottom panels).  We searched for periods or quasi-periodic oscillations in both the {\it Chandra} and {\it XMM-Newton} lightcurves, using {\it powspec}, {\it efsearch} and {\it efold}, but none of the signal peaks is significant above the noise level.
 In particular, we searched for periods $\sim$1 s in the 2012 and 2017 EPIC-pn lightcurves (binned to 0.1 s), by analogy with typical periods found in ULX pulsars ({\it e.g.}, \citealt{bachetti14,sathyaprakash19,castillo19}). For this, we used the {\it LombScargle} routine in {\sc astropy} version 3.2.1 \citep{astropy18}. For the 2012 lightcurve, we found a probability $>$40\% that any of the peaks in the Lomb-Scargle periodogram are due to random fluctuations of photon counts; for the 2017 dataset, that probability is $>$50\%. Thus, we cannot detect any significant period in this source.
The relatively low number of counts and signal-to-noise ratio of these observations are not sufficient for any more detailed variability analysis on such short timescales.

\subsection{Spectral Properties}

We have enough counts for detailed modelling of the spectra from {\it XMM-Newton} in 2012 and 2017, and {\it Chandra} in 2016.  As a first step, we tried a simple power-law model ({\it tbabs} $\times$  {\it tbabs} $\times$ {\it pow}) with a fixed  Galactic absorption component $N_{\rm {H,Gal}} = 2 \times 10^{21}$ cm$^{-2}$ \citep{kalberla05} and a free intrinsic component. The model provides a good fit for the 2016 spectrum (partly because of the more limited band coverage of ACIS) but leaves significant systematic residuals in the 2012 and 2017 spectra. One finding that is already obvious from a power-law fit is that the 2017 spectrum is significantly softer ($\Gamma \approx 3.5$) than the other two. Another finding is a strong residual feature consistent with an unresolved emission line at $E \approx 0.66$ keV in the 2017 and (at a weaker level)  in the 2012 data. The presence of this feature is significant for every choice of continuum model (power-law, and every other continuum component tested in the rest of this section); we will discuss this line in more details later.

Next, we tried another simple one-component model: {\it tbabs} $\times$ {\it tbabs} $\times$ {\it diskbb}. The disk-blackbody model does not improve on the power-law model for any of the three spectra: in simple terms, it is too curved for the observed datapoints. We improved on the disk-blackbody model in two alternative ways: by adding a second (softer) thermal component ({\it tbabs} $\times$ {\it tbabs} $\times$ ({\it diskbb} $+$ {\it bbodyrad})), and by using a $p$-free disk model ({\it tbabs} $\times$ {\it tbabs} $\times$ {\it diskpbb}), with $p<0.75$.

The double-thermal model provides a good fit ($\chi^2_{\nu} < 1.2$) at all three epochs (Table 2). The lower-temperature blackbody has a characteristic temperature $kT_{\rm bb} \approx 0.15$--0.3 keV, typical of the soft excess observed in many ULXs \citep{kajava09,gladstone09,kaaret17,zhou19}, and a characteristic radius of $\sim$10$^3$ km  for all three epochs (again, a common feature for this class of systems). The best-fitting inner disk  color temperatures in 2012 ($kT_{\rm in} \approx 1.25$ keV) and 2016 ($kT_{\rm in} \approx 1.14$ keV) are consistent with the temperature expected from the inner disk of a stellar-mass BH at the Eddington limit. The physical inner-disk radius $R_{\rm in}$ is assumed to be the innermost stable circular orbit, and is defined as $R_{\rm in} \approx 1.19 r_{\rm in}$, where $r_{\rm in}$ is the best-fitting radius in the {\it diskbb} model in {\sc xspec} \citep{kubota98};  the standard scaling factor 1.19 accounts for the hardening factor ($\approx$1.7) and the fact that the peak effective temperature occurs slightly outwards of the inner disk radius, at $R \approx (49/36) R_{\rm in}$ \citep{kubota98}. In our case, we find $R_{\rm in} \sqrt{\cos \theta} \approx 55$ km in 2012, and $R_{\rm in} \sqrt{\cos \theta} \approx 70$ km in 2016; again, these values are typical of stellar-mass BHs at high luminosities. Instead, the best-fitting temperature $kT_{\rm in} \approx 0.7$ keV, determined for the 2017 spectrum, is too low for the observed luminosity.  For comparison, in other BH X-ray binaries in which the accretion disk emission is still the dominant component of the X-ray spectrum at $L_{\rm X} \approx L_{\rm {Edd}} \approx 10^{39}$ erg s$^{-1}$, the peak colour temperature $kT_{\rm {in}} \approx 1.2$--1.5 keV  \citep{k04,s07,sutton13}.  This suggests that although the curvature of a disk model provides a formally good fit to the data, it does not provide the correct physical interpretation of the emission at that epoch. The de-absorbed 0.3--10 keV luminosity, defined as $4\pi d^2$ times the de-absorbed flux $f$ \footnote{The luminosity $L$ of a disk emission component is more properly defined as $L = 4 \pi d^2 f/\cos \theta$; however, it is also customary to adopt $L = 2\pi d^2 f$ in the absence of information about the viewing angle $\theta$.}, is $L_{0.3-10} \approx 2 \times 10^{39}$ erg s$^{-1}$, both in 2012 and 2016 (Table 2). 

The $p$-free disk model is a simple approximation to slim disk models \citep{abramowicz88,watarai01}, suitable for stellar-mass BHs near or slightly above their Eddington limit. The definition of the parameter $p$ is that the effective temperature on the disk scales as $T(R) \propto R^{-p}$; the standard disk case is $p=0.75$, while near-Eddington BHs tend to have $p \lesssim 0.6$ \citep{sutton17}. In our case, the $p$-free disk provides statistically equivalent fits to the double-thermal model, for the 2012 and 2016 spectra ($\chi^2_{\nu} \approx 1.1$ at both epochs). The best-fitting temperature $kT_{\rm in} \approx 1.5$ keV in 2012 and $kT_{\rm in} \approx 1.4$ keV in 2016, consistent with typical values expected for slim disks near the Eddington limit. For a $p$-free disk, the physical inner radius is usually defined as $R_{\rm in} \approx 3.19 r_{\rm in}$ \citep{vierdayanti08}; in our case, we obtain $R_{\rm in} \sqrt{\cos \theta} \approx 75$ km both in 2012 and in 2016. De-absorbed luminosities are $\approx$1.5--2 $\times 10^{39}$ erg s$^{-1}$ in both epochs. Instead, a $p$-free disk is not a good fit for the 2017 spectrum ($\chi^2_{\nu} \approx 1.6$); specifically, it does not model well the characteristic downturn seen in the data above 2 keV. 

For our fourth attempt to model the data, we used a comptonization model: {\it tbabs} $\times$ {\it tbabs} $\times$ ({\it bbodyrad} $+$ {\it comptt}). Here, the {\it bbodyrad} component plays the role of seed thermal emission. Replacing {\it bbodyrad} with a {\it diskbb} seed component does not change our results, because we are only looking at the Wien section of the curve. This model provides a good fit for all three epochs. For 2012 and 2016, it is statistically equivalent to the double thermal model and the $p$-free model; for 2017, it is as good as the double thermal model.
The seed photon temperature is $kT_0 \approx 0.15$ keV in 2012 and 2017, while it is unconstrained ($\lesssim 0.1$ keV) in 2016. The electron temperature in the Comptonizing cloud, which determines the location of the high-energy downturn, is $\sim$1 keV in 2012 and 2017, while it is unconstrained in 2016, where the presence of a high-energy downturn is not statistically significant (mostly because of the more limited band coverage in {\it Chandra}). Low temperatures and high optical depths are one of the defining properties of ULXs \citep{gladstone09}, when their spectra are fitted with Comptonization models. The most common interpretation for this kind of spectra is that the harder photons from the inner disk region are downscattered in a thick disk outflow, seen at high inclination angles. The de-absorbed luminosity for the Comptonization model in 2012 and 2017 is $\approx$1.5--3 $\times 10^{39}$ erg s$^{-1}$, similar to the luminosity inferred from the other models. Instead, the luminosity is unconstrained in the 2016 {\it Chandra} spectrum because the temperature of the seed thermal component is too low and its normalization is unconstrained; if we neglect the contribution of the seed photons, the luminosity is $\approx$4 $\times 10^{39}$ erg s$^{-1}$ in 2016.  Alternatively, we fixed the temperature of the blackbody and seed thermal components at $kT_0 = kT_{\rm bb} \equiv 0.10$ keV, to reduce the degeneracy between temperature, normalization, and absorption column. The best-fitting parameters are essentially unchanged; the de-absorbed luminosity has a best-fitting value of $\approx$5.4 $\times 10^{39}$ erg s$^{-1}$ (of which, $\approx$3.5 $\times 10^{39}$ erg s$^{-1}$ from the Comptonized component), with a 90\% lower limit of $\approx$2.6 $\times 10^{39}$ erg s$^{-1}$ and a badly constrained upper limit of $\approx$3.3 $\times 10^{40}$ erg s$^{-1}$.

In summary, all three spectra are consistent with a mildly super-Eddington stellar-mass BH. The EPIC spectrum from 2012 is more consistent with a broadened disk regime \citep{sutton13,gladstone09} but can also be the result of Comptonization. The 2016 ACIS spectrum is softer than in 2012, but cannot be reliably classified because of the limited band coverage. The 2017 EPIC spectrum is significantly softer than the other two, is best modelled with a Comptonization model, and belongs to the soft ultraluminous regime. The de-absorbed luminosity was $\approx$1.5--2$\times 10^{39}$ erg s$^{-1}$ in 2012, $\approx$2--4$\times 10^{39}$ erg s$^{-1}$ in 2016, and $\approx$2--3$\times 10^{39}$ erg s$^{-1}$ in 2017. The softness of the 2017 spectrum compared with the other two is obvious when we plot unfolded spectra (Figure~\ref{ufs}), based on Comptonization models for consistency.

For all continuum spectral models described above, we found a significant emission line residual at $E = 0.66 \pm 0.01$ keV in the 2017 spectrum (Table 2); a similar residual but with much lower significance is also seen in the 2012 spectrum. We modelled the residual with a Gaussian emission line: we found an equivalent width of $95^{+55}_{-60}$ eV in 2017 (almost independent of the choice of continuum model) and an intrinsic line luminosity of $\approx$2.4 $\times 10^{38}$ erg s$^{-1}$, that is $\approx$10 per cent of the total intrinsic luminosity in the EPIC band.  The full width at half maximum of the line ($\sigma_{\rm {line}}$ in Table 2) is consistent with zero (that is, the intrinsic velocity broadening of the line is much less than the instrumental broadening), and constrained to be $\lesssim$30 eV ($\approx$14,000 km s$^{-1}$) to the 90\% confidence level in the 2017 spectrum, with only small differences between the different continuum models.

To assess the significance of the line component, first of all we note that the 90\% lower limit of the line normalization parameter (photon flux) is $>0$ both in the 2012 and 2017 spectral fits (Table 2). The improvement in the fit statistic when we include the Gaussian emission line component is $\Delta \chi ^2 >11.4$ for all the continuum spectral models. We obtained a more rigorous statistical constraint with the likelihood ratio test {\it lrt} in {\sc xspec}: we ran 10000 simulations for various pairs of models with and without the line component. The significance of the line in 2012 is only $\approx$70\%, while in 2017 it is $>$99.8\% regardless of the continuum model. In fact, Figure~\ref{spec} shows that the 0.66 keV emission line in the 2017 spectrum is evident even simply by eye. We also built and inspected EPIC-pn and MOS images in the 0.60--0.70 keV band (Figure~\ref{oxygen}) for the 2017 dataset, to make sure that the emission is dominated by the point-like ULX and is not contaminated for example by diffuse hot gas in a spiral arm. We compared the narrow-band images from 2017 with those from earlier epochs, and from {\it Chandra} when the ULX was not detected, to exclude the possibility of a pre-existing supernova remnant at that location. As a result of these tests, we are confident that the $\gtrsim$10$^{38}$ erg s$^{-1}$ line emission seen does come from the ULX. 

The most likely identification is the O {\footnotesize{VIII}} Ly$\alpha$ line (rest frame energy of 0.654 keV). Such line was strongly detected in other soft ULXs, and is interpreted as a signature of fast outflows \citep{pinto16,pinto17,kosec18}.  Although it is not a surprise to detect this line in a ULX, it is unusual to see it so strongly and significant even in a moderate-quality spectrum at CCD resolution (see also Section 4.2 for a comparison with other ULXs).

 Another line that has been found associated with ULX outflows is the O {\footnotesize{VII}} triplet at 0.56 keV \citep{pinto16,pinto17,kosec18}. In our spectra, we do not have enough counts to detect or place meaningful constraints on the presence of this line. The main reason for the lower signal-to-noise ratio is that photoelectric absorption is stronger at 0.56 keV than at 0.65 keV, and, conversely, the pn and MOS effective areas are lower. Nonetheless, we tried adding a narrow line with energy fixed at 0.56 keV and estimated the 90\% upper limit to its normalization, for each of the four spectral models described before. The result of this test is that a 0.56 keV with the same photon flux as the 0.66 keV line ({\it i.e.}, $\sim$10$^{-5}$ photons cm$^{-2}$ s$^{-1}$, or $EW\lesssim$75 eV) is still consistent with our data, even though this additional line would not stand out ``by eye'' in the spectra. The lack of strong constraints on the strength of the O {\footnotesize{VII}} triplet means that we cannot exclude a (minor) contribution to the 0.66 keV line from the O {\footnotesize{VII}} He$\beta$ line at 0.67 keV. 


The 2017 EPIC spectrum shows also hints of other emission-line features in the soft band (Figure~\ref{spec}, bottom right panel); for example the Mg {\footnotesize{XI}} triplet at 1.35 keV.  The presence of other possible line features motivated us to try and replace the single Gaussian component in our 2017 spectral fits with a thermal plasma component, {\it vapec} in {\sc xspec}, suitable for collisionally ionized gas. Keeping the metal abundances of all elements at the solar value does not provide an improvement compared with the simpler model fits without a Gaussian line. Similarly, leaving all abundances free but locked together does not improve the fit. Instead, we found that we can obtain good fits (Table 3) by leaving the abundance of some elements (O, Ne, Mg and Si) free and keeping the other abundances at the solar level. The very high relative abundances of the $\alpha$ elements required to fit our spectrum may be a clue for the type of donor star, as we shall discuss later (Section 4.3). The de-absorbed luminosity in the {\it vapec} component is $\approx$3 $\times 10^{38}$ erg s$^{-1}$. 

For the 2017 spectrum, we then replaced the solar-metallicity intrinsic absorption component ({\it tbabs}) with  variable abundance absorption models (we tried {\it tbvarabs} and {\it vphabs}). We did not find any significant improvement or any change in the emission line properties when the oxygen abundance in the intrinsic absorber is a free parameter. In fact, the oxygen abundance in the intrinsic absorber is constrained to be $\lesssim$ 2 times solar at the 90\% confidence level. This suggests that the gas responsible for the intrinsic absorption is not the same oxygen-rich gas responsible for the strong 0.66 keV emission line; the intrinsic absorption component may be located in the halo and disk of NGC\,6946, rather than within the binary system.

Finally, we modelled the stacked spectra from the {\it Swift}/XRT observations in 2017, 2018 and 2019. Because of the low number of counts, the detailed models described above are degenerate and cannot be adequately constrained; however, we can still get useful information on the hardness evolution. To do so, we fixed the intrinsic column density $N_{\rm H} = 2 \times 10^{21}$ cm$^{-2}$, as derived from most of the {\it Chandra} and {\it XMM-Newton} spectral fits, and fitted the power-law slope and normalization. We found that the ULX spectrum has hardened in recent years: from $\Gamma = 3.2 \pm 0.5$ in 2017 (consistent with the contemporaneous {\it XMM-Newton} observations), to a more moderate $\Gamma = 2.1 \pm 0.4$ in 2018, and $\Gamma = 2.6 \pm 0.5$ in 2019. At the same time, the X-ray luminosity has remained approximately constant at $\approx$3 $\times 10^{39}$ erg s$^{-1}$.

\subsection{Constraints on the optical counterpart}

We have already described (Section 3.1) how we used the relative offsets to MF16 and SN\,20017eaw to pinpoint the location of the X-ray transient in the {\it HST} images. We detect only one faint optical source inside the error circle (Figure~\ref{counterparts}), in the 2016 ACS images, both in the F606W and F814W bands. The same source is too faint to be detectable in the 2004 and 2017 observations (shorter exposure times). Thus, we cannot constrain the source variability.  Aside from the positional coincidence, there is no direct evidence that this optical source is the actual donor star for the transient X-ray source. 

We measured an apparent brightness $m_{\rm{F606W}} = (26.35 \pm 0.15)$ mag, and $m_{\rm{F814W}} = (26.05 \pm 0.15)$ mag (corrected to infinite aperture). After correcting for line-of-sight extinction ($A_{\rm{F606W}} = 0.85$ mag, $A_{\rm{F814W}} = 0.52$ mag: \citealt{2011ApJ...737..103S}), we obtain absolute magnitudes $M_{\rm{F606W}} = (-3.95 \pm 0.15)$ mag, and $M_{\rm{F814W}} = (-3.90 \pm 0.15)$ mag ( Table 4). This is consistent with an early B star (main sequence or subgiant), while it rules out O stars and supergiants.  It is a type of optical counterpart very common in ULXs \citep{gladstone13,motch14}. However, in Section 4.3 we will propose a white dwarf scenario for the donor star, in which case this faint optical source is unrelated to the ULX.

\section{Discussion}

\subsection{Transient behaviour}

There are at least two types of transient X-ray sources that reach ultraluminous regime. One type, well exemplified by CXOU J203451.1$+$601043, turns on as a ULX after years of non-detection (typically at least two orders of magnitude fainter), and then remains in the super-Eddington regime for many years. Another example of this behaviour is the transient ULX in M\,83 \citep{soria12,soria15}. In our galaxy, GRS\,1915$+$105 turned on in 1992 \citep{castro-tirado92} and has been near the Eddington limit ever since, at least until mid-2018 when it started to decline towards the low/hard state \citep{negoro18,miller19,rodriguez19}. The second type of ultraluminous transient is well represented by another source recently discovered in NGC\,6946, labelled ULX-4 in \cite{earnshaw19b}; its outburst lasted only $\sim$10 d. Another example of short-duration ULX transient is the first ULX discovered in M\,31, which went back to quiescence after a few months \citep{middleton13} . In the Milky Way, V404 Cyg provides the most notable example of short-duration outburst reaching the Eddington luminosity, and then declining to quiescence, between 2015 June and August \citep{motta17,munoz-darias16,kimura16,sivakoff15}.

The most commonly accepted theoretical interpretation of X-ray outbursts in Galactic X-ray binaries is based on the thermal-viscous disk instability \citep{king98,dubus01,lasota01}, which depends on the existence of an outer disk region where hydrogen is mostly neutral. The natural extension of this model for long-duration super-Eddington transients (such as CXOU J203451.1$+$601043) requires the simultaneous presence of two ingredients: a sufficiently high irradiation of the outer disk to keep it in the ionized state, and a sufficiently high accretion rate to keep the source at or above the Eddington limit for years. If the donor is a low-mass star, which cannot keep a persistently high mass transfer rate through the L1 Lagrangian point, a long-duration phase of super-Eddington accretion may still be achieved if the disk is very large\footnote{ A large disk is probably a necessary but not a sufficient condition for a long outburst: for example, the transient Galactic stellar-mass BH V404 Cygni, mentioned before, also has a large disk despite its short-duration outburst. This has been interpreted \citep{munoz-darias16} as the effect of a strong outflow that disrupted the supply of accreting matter from the outer disk to the inner disk.}
and able to store enough mass prior to an outburst (the mass stored in the disk is proportional to $R_{\rm out}^3$, and the peak luminosity scales as $R_{\rm out}^2$: \citealt{king98}). For example, in the case of GRS\,1915$+$105, the estimated disk size is $\sim$10$^{12}$ cm and the mass in the disk is $\sim$10$^{28}$ g, which suffices to keep the X-ray source at or above 10$^{39}$ erg s$^{-1}$ for decades \citep{done04}.
As for the other requirement ({\it i.e.}, a strong irradiation of the outer disk), broadband spectral modelling of observational data suggests \citep{sutton14} reprocessing fractions $\sim$10$^{-3}$ for sub-Eddington stellar-mass X-ray binaries and some ULXs, and $\sim$10$^{-2}$ for other ULXs, specifically those with soft X-ray spectra. The reason for the enhanced irradiation factor in some ULXs may be that their strong disk wind intercepts and scatters a fraction of photons emitted in the polar funnel and re-directs them onto the outer disk \citep{sutton14}. 

This is not the only viable explanation for transient behaviour in ULXs. If a ULX is powered by a NS rather than a BH, transitions between the propeller and the accretor state will cause a transient behaviour \citep{dallosso15,tsygankov16,earnshaw18}. The drop in luminosity happens when the magnetospheric radius of the NS (the radius at which the magnetic pressure ``stops'' the inflowing matter) becomes larger than the corotation radius of the disk, thus creating a centrifugal barrier \citep{illarionov75,stella86}. The location of the magnetospheric radius depends on the NS spin, the mass accretion rate, and the strength of the NS magnetic field. The existence of a luminosity gap in the long-term light curve of a transient ULX, between $\sim$a few 10$^{37}$ erg s$^{-1}$ and $\sim$10$^{39}$ erg s$^{-1}$, would be a clue in favor of an accretor/propeller switch model, rather than a thermal-viscous disk instability (where we expect the system to evolve smoothly between higher and lower luminosities without discrete jumps). For CXOU J203451.1$+$601043, all we can say at the moment is that the system was never detected at any other luminosity $<$10$^{39}$ erg s$^{-1}$;  thus, the accretor/propeller model is still viable. Future observations of CXOU J203451.1$+$601043 will be needed to determine whether, when and how it will decline.

\subsection{The soft ultraluminous regime and the oxygen line}

We showed (Section 3.3) that the EPIC spectrum from 2012 is consistent either with a broadened disk regime or Comptonization, while the 2017 spectrum is significantly softer, not consistent with disk models, and suggests down-scattering of the direct X-ray emission in a cooler medium such as the disk outflow. A high-energy downturn already at a photon energy $\approx$3 keV (or, equivalently, a steep photon index $\Gamma \approx 3.5$ when fitted in the 0.3--8 keV band) make the 2017 state of this ULX even softer than NGC\,5408 X-1, {\it i.e.}, the source usually taken as a standard for the soft ultraluminous regime  \citep{sutton13,middleton14}. The 2017 spectrum puts it in the same class as NGC\,55 X-1 \citep{stobbart04,pinto17} and NGC\,247 X-1 \citep{feng16}, {\it i.e.} the two sources that are in a transitional state between the classical ULX regime and the supersoft regime. The very soft spectral appearance found in 2017 is also reminiscent of the eclipsing ULX CXOM51 J132940.0$+$471237 in M\,51 \citep{urquhart16}. The high scattering optical depth ($\tau \approx 13$) fitted to the {\it comptt} model is consistent with the relation between optical depth and coronal temperature found by \cite{pintore14}, again at the very soft end of the sequence.

The spectral and timing properties of the soft ultraluminous regime are generally attributed to our viewing angle passing through the thick disk outflow \citep{kawashima12,sutton13,pintore14,middleton15a,pinto17,narayan17}. This reduces or suppresses our detection of hard X-ray photons (directly emitted from the inner disk region), and increases the contribution of down-scattered soft X-ray photons. Spectral softening may be caused by an increase of the radiatively driven wind mass-loss rate and therefore of its optical depth, which is a function for example of the accretion rate \citep{poutanen07}. In the case of CXOU J203451.1$+$601043, the de-absorbed luminosity fitted to the observed spectral datapoints varies only by a factor of 2 between the {\it Chandra}, {\it XMM-Newton} and {\it Swift} observations; however, those luminosity estimates correct only for the effect of cold absorption, not for the down-scattering of hard X-ray photons into the soft X-ray or far-UV band. It is plausible that the intrinsic luminosity in 2017 would be much higher, if we could see the direct emission unaffected by the down-scattering outflow (for example, if we had a pole-on view). Precession of the viewing angle may also change the scattering optical depth seen by distant observers, for constant wind properties.

If our interpretation of CXOU J203451.1$+$601043 as an extreme example of the soft ultraluminous regime is correct, we expect two other properties associated with this regime. One is the large amount of short-term variability \citep{middleton15a}. Moderate intra-observational variability is observed in our X-ray timing analysis; however, the observed count rate is too low to constrain the root-mean-square fractional variability at high frequencies. Thus, we cannot make any firm conclusions on whether this source is more variable than other ULXs.

The second property that should be associated with a thick down-scattering outflow is the presence of spectral residuals in the soft X-ray band, caused by blends of emission and absorption lines \citep{middleton15b,pinto16,pinto17}. Indeed, we have shown the presence of at least one strong O {\footnotesize{VIII}} emission line at $E = (0.66 \pm 0.01)$ in the 2017 EPIC spectrum, with a luminosity $\approx$2 $\times 10^{38}$ erg s$^{-1}$ and an equivalent width of $\approx$100 eV. The line was significantly stronger than in the 2012 spectrum. We have also shown hints of other likely residuals around 
1.35 keV (Mg {\footnotesize{XI}}) and 1.7 keV (Mg {\footnotesize{XII}}). The  luminosity of the O {\footnotesize{VIII}} emission line in 2017 was an order of magnitude higher than that of  analogous lines detected in other ULXs such as NGC\,1313 X-1 and NGC\,5408 X-1 \citep{pinto16,middleton15b},  and also an order of magnitude higher than that of the  O {\footnotesize{VIII}} line in NGC\,55 X-1 \citep{pinto17}, despite the similarity in the X-ray continuum.

\subsection{An ultracompact ULX?}

Why is the oxygen line so strong? We speculate that the donor star is oxygen-rich, well above solar abundance. One scenario is that the donor star is an oxygen-rich Wolf-Rayet (WO subclass; \citealt{crowther07,mcclelland16}). The optical luminosity of a WO star is low enough \citep{sander19} to be consistent with our upper limit of $\approx$ $-4$ mag for the optical counterpart.  Our spectral modeling suggests that the source is likely viewed at a high inclination angle. Consequently, if the donor star was a Wolf-Rayet, we would expect to see strong sinusoidal variability or even eclipsing behaviour in the X-ray flux, by analogy with other Wolf-Rayet X-ray binaries \citep{qiu19a,qiu19b}. Instead, no such variability is detected in the lightcurves of this ULX (Figure 5). A more plausible scenario is that the donor star is a CO or (preferably) an O-Ne-Mg white dwarf; the latter are the most massive sub-class of white dwarfs \citep{truran86,shara94}, formed from B-type stars with initial masses just below the limit ($\approx 8 M_{\odot}$) for supernova explosions. In order to form a luminous X-ray binary, the white dwarf must be filling its Roche lobe in an ultracompact system ({\it e.g.}, \citealt{vanhaaften12}). Strong,  relativistically broadened O {\footnotesize{VIII}} Ly$\alpha$ lines attributed to Compton reflection have been seen from some (sub-Eddington) ultracompact X-ray binaries (UCXBs) in the Milky Way \citep{madej14,madej11,madej10}.  The origin of the emission line in CXOU J203451.1$+$601043 may be different (an outflow rather than Compton reflection on the inner disk surface), but we mention this analogy simply as an indication of an oxygen-rich accretion flow. It is also theoretically possible that UCXBs reach super-Eddington luminosities: ULXs in globular clusters, such as the one in the RZ 2109 cluster of NGC\,4472 \citep{maccarone07}, have been interpreted as UCXBs. In particular, a strong [O {\footnotesize{III}}] $\lambda$5007 emission line was detected in the optical spectra of the RZ 2109 source \citep{steele14}, and was interpreted as evidence of a hydrogen-poor, oxygen rich donor, and of an outflow powered by the accreting compact object. In the soft X-ray band, several {\it XMM-Newton} and {\it Chandra} spectra of the UCXB in RX 2109, presented by \cite{dage18}, also show emission residuals at 0.6--0.7 keV consistent with O {\footnotesize{VIII}} Ly$\alpha$ emission. 

A possible  issue with the UCXB scenario is that the candidate ULX UCXBs suggested in the literature so far are all in globular clusters, where dynamical formation is strongly enhanced; instead, CXOU J203451.1$+$601043 is in the field, in or near a spiral arm,  almost certainly not in a globular cluster, because its optical counterpart is fainter than $M_I \approx -4$ mag. The typical optical luminosity distribution of old globular clusters spans between $-11 \lesssim M_I{\mathrm{(mag)}} \lesssim -5$ ({\it e.g.}, \citealt{jordan07,barmby00,secker92}). On the other hand, several Galactic UCXBs are also found in the field, outside globular clusters or the bulge \citep{cartwright13}. The transient nature of CXOU J203451.1$+$601043 is also a puzzle: the disk in an ultraluminous UCXB is too hot (and therefore fully ionized) to undergo thermal-viscous instabilities; instead, the transient behaviour could be caused by mass transfer instabilities from the donor star, or an accretor/propeller switch if the compact object is a magnetized NS.  




 
\section{Conclusions and future prospects}

Using archival {\it Chandra} and {\it XMM-Newton} observations, we have identified a previously unrecognized, transient ULX in NGC\,6946. The source was undetected, at luminosities $\lesssim$ a few $10^{37}$ erg s$^{-1}$, in all observations until 2008, and always detected at luminosities $\approx$ 1.5--3 $\times 10^{39}$ erg s$^{-1}$ in all observations between 2012 and 2019. We pointed out a few interesting properties that help our understanding of the ultraluminous regime. First, we showed that the source is extremely soft: if modelled with a Comptonization spectrum, the electron temperature is $\approx$0.7 keV in the 2017 EPIC spectrum.  Spectral evolution between different epochs suggests a change in the down-scattering wind properties. In its softest state (2017 {\it XMM-Newton} observations), CXOU J203451.1$+$601043 is in the transitional regime between standard ULXs and ultraluminous supersoft sources. This finding supports the argument that soft ULXs and ultraluminous supersoft sources are fundamentally similar systems, distinguished by the optical depth of the scattering wind along our line of sight. Second, we showed that CXOU J203451.1$+$601043 has another property associated with super-Eddington outflows: strong line residuals in the soft X-ray band. In particular, the strong 0.66 keV emission line (likely to be the O {\footnotesize{VIII}} Ly$\alpha$ line) is the most outstanding feature of this source in its softest state. Very few ULXs display wind emission lines so strong that can be easily identified and modelled even at CCD resolution. We speculate that the strong oxygen line is evidence of an oxygen-rich donor star, such as an O-Ne-Mg white dwarf. If so, it would be the first example of an ultracompact ULX outside a globular cluster, adding more variety of formation channels to the already heterogeneous ULX population. 

Future follow-up studies of CXOU J203451.1$+$601043 may provide important constraints on at least three unsolved problems. First, it will be important to monitor the duration of the super-Eddington regime, which has already lasted at least 7 years and shows no sign of decline. When (if) the source does decline, the crucial test will be whether the outburst decline follows the characteristic hardness-luminosity tracks of stellar-mass BHs \citep{fender04} or instead shows a sudden disappearance expected for accretor/propeller transitions in NSs. Second, future observations of this ULX offer a chance to determine a quantitative relation between the strength of the 0.66 keV emission line (and of other line residuals), and the energy of the downturn in the continuum, linking the imprint of the wind on line and continuum emission. Third, we speculate that if CXOU J203451.1$+$601043 is an UCXB,  X-ray lightcurves may reveal a characteristic period $\sim$10 min, but we may have to wait until deeper observations with {\it Athena} to find out.

\vspace{0.5cm}

We thank the referee for her/his careful reading and useful suggestions. We are grateful to Song Wang (NAOC) for assistance with the timing analysis and Hua Feng (Tsinghua University) for helpful discussion. JW was supported by the National Key R\&D Program of China (2016YFA0400702) and the National Science Foundation of China (U1831205, 11473021, 11522323). For this research, we used data obtained from the Chandra Data Archive, and {\sc ciao} software provided by the Chandra X-ray Center. We also used data obtained with {\it XMM-Newton}, an ESA science mission with instruments and contributions directly funded by ESA Member States and NASA. Our optical results were based on images from the NASA/ESA {\it Hubble Space Telescope}, obtained from the Space Telescope Science Institute, which is operated by the Association of Universities for Research in Astronomy, Inc., under NASA contract NAS 5-26555.

\clearpage

\begin{table*} 
\caption{{\it Chandra}, {\it XMM-Newton} and {\it Swift} observations of NGC\,6946, and luminosity of CXOU J203451.1$+$601043}  
\begin{center}  
\begin{tabular}{lcccr}  
\hline \hline\\[-3pt]    
ObsID &  Good Time Interval & Observation Date & Observed X-ray Flux$^a$  & X-ray Luminosity$^a$\\
 &  (ks)  &  &  (erg cm$^{-2}$ s$^{-1}$) &  (erg s$^{-1}$)\\[5pt]
\hline  \\[-3pt]
\multicolumn{5}{c}{{\it Chandra}/ACIS} \\ [5pt]
\hline \\[-3pt]

1043 &58.3 &2001-09-07 & $<1.2 \times 10^{-15}$ & $< 2.1 \times 10^{37}$ \\ [3pt]
4404 &28.7 &2002-11-25 & $<1.4 \times 10^{-15}$  &  $< 2.5 \times 10^{37} $ \\ [3pt]
\hline \\[-3pt]

4631 &28.4 &2004-10-22 &   \multirow{3}[2]{*}{$<1.1 \times 10^{-15}$} 
& \multirow{3}[2]{*}{$ < 2.0 \times 10^{37}$}  \\[3pt]
4632 &25.2 &2004-11-06 &  & \\[3pt]
4633& 26.6 &2004-12-03 &  & \\[3pt]
\hline\\[-3pt]
13435    &  20.4 & 2012-05-21 & $4.0^{+0.6}_{-0.6} \times 10^{-14}$ & $1.1^{+0.2}_{-0.2} \times 10^{39}$ \\[3pt]
\hline\\[-3pt]
17878 &  40.0 & 2016-09-28  & \multirow{2}[2]{*}{$2.7^{+0.2}_{-0.2} \times 10^{-13}$}   & \multirow{2}[2]{*}{$4.8^{+0.4}_{-0.4} \times 10^{39}$} \\[3pt]
19887 & 18.5 &  2016-09-28  & & \\[3pt]
\hline\\[-3pt]
19040   & 9.8 &  2017-06-11  & $0.6^{+0.1}_{-0.1} \times 10^{-14}$ & $1.7^{+0.3}_{-0.3} \times 10^{39}$ \\[3pt]
\hline \\[-3pt]
\multicolumn{5}{c}{{\it XMM-Newton}/EPIC} \\ [5pt]
\hline \\[-3pt]

   	0093641501      &   	0.6   &  	2003-04-18 & \multirow{3}[2]{*}{$ < 1 \times 10^{-14}$} & \multirow{3}[2]{*}{$ < 2 \times 10^{38}$} \\[3pt] 
      	0093641601   &   2.2 &      2003-05-17&  &  \\[3pt]
   0093641701	     &  1.2    &   		2003-06-18   &  & \\[3pt]
\hline\\[-3pt]
    	0200670101      &      3.9        &         	2004-06-09     & \multirow{4}[2]{*}{$ < 3.1 \times 10^{-15}$} & \multirow{4}[2]{*}{$ < 3 \times 10^{37}$}     \\[3pt]
     	0200670201     &   12.7 &        	2004-06-11      &  & \\[3pt]
     	0200670301     &      	11.3         &       	2004-06-13    &  &      \\[3pt]
     0200670401     &           8.8   &     2004-06-25   &  & \\[3pt]
\hline \\[-3pt]
       0401360101   &  18.7   &        	2006-05-23  &     \multirow{3}[2]{*}{$ < 2 \times 10^{-15}$} & \multirow{3}[2]{*}{$ < 1.5 \times 10^{37}$}      \\[3pt]
          	0401360201 &         4.7      &      2006-06-02       &    &   \\[3pt]
0401360301 &  4.9 &      	2006-06-18  & &    \\[3pt]
\hline \\[-3pt]
 0500730101           &       	26.0         &   2007-11-08   & \multirow{2}[2]{*}{$ < 1.2 \times 10^{-15}$}   &   \multirow{2}[2]{*}{$ < 2 \times 10^{37}$}   \\ [3pt]
0500730201 &         	31.7   &          2007-11-02   &    &    \\[3pt]
\hline \\[-3pt]
    0691570101      &      109.3          &          2012-10-21      &  
    $1.6^{+0.1}_{-0.1} \times 10^{-13}$   & $1.6^{+0.1}_{-0.1} \times 10^{39}$   \\[3pt]
\hline \\[-3pt]
     0794581201     &          	43.1     &      2017-06-01       &  
     $0.6^{+0.1}_{-0.1} \times 10^{-13}$ & $2.8^{+2.0}_{-0.1} \times 10^{39}$   \\ [3pt]  
\hline   \\[-3pt]
\multicolumn{5}{c}{{\it Swift}/XRT} \\ [5pt]
\hline \\[-3pt]

   		31113001 to 31113004      &   10   &  	2008-02-04 to 2008-02-14 & $ < 2 \times 10^{-14}$ & $ < 4 \times 10^{38}$ \\[3pt] 
   		49820001 to  49820003     &   7   &  	2013-05-31 to 2013-06-04 & $1.4^{+0.4}_{-0.4} \times 10^{-13}$ & $2.4^{+0.5}_{-0.5} \times 10^{39}$ \\[3pt] 
   		10130001 to 10130029      &   44   &  	2017-05-13 to 2017-09-17 & $1.1^{+0.2}_{-0.2} \times 10^{-13}$ & $3.6^{+0.4}_{-0.4} \times 10^{39}$ \\[3pt] 
   		94059001 to 94059044      &   43   &  	2018-04-01 to 2018-12-27 & $2.2^{+0.3}_{-0.3} \times 10^{-13}$ & $2.8^{+0.4}_{-0.4} \times 10^{39}$ \\[3pt] 
   		94059045 to 94059072      &   17   &  	2019-01-06 to 2019-04-06 & $1.7^{+0.3}_{-0.3} \times 10^{-13}$ & $3.2^{+0.4}_{-0.4} \times 10^{39}$ \\[3pt] 
\hline\\[-3pt]
\end{tabular} 
\label{tab1}
\end{center}
\begin{flushleft} 
\footnotesize{
$^a$: for the {\it Chandra} observations, we estimated observed fluxes and intrinsic luminosities (or their respective upper limits) in the 0.3--10 keV band with the {\sc ciao} task {\it srcflux}, assuming a power-law model with photon index $\Gamma = 2.5$ and total column density $N_{\rm H} = 4 \times 10^{21}$ cm$^{-2}$ (twice the Galactic line-of-sight value). This model was chosen because it approximates the best-fitting power-law model for the 2016 {\it Chandra} spectrum (Table 2). For {\it XMM-Newton} observations, we also used a simple power-law model to convert from count rates to fluxes and luminosities in this Table. When we had enough counts for a detailed fit (2012 and 2017 {\it XMM-Newton} observations), we used the best-fitting values (Table 2); for the non-detections, we assumed $\Gamma = 2.5$ and $N_{\rm H} = 4 \times 10^{21}$ cm$^{-2}$. For {\it Swift}, we fixed $N_{\rm H} = 4 \times 10^{21}$ cm$^{-2}$,  then used {\sc pimms} to estimate the power-law photon index that best approximates the observed (1.5--10)/(0.3--1.5) hardness ratio of each observation. We used those indices to infer fluxes and luminosities of the respective observations; when not enough counts are available, we assumed again $\Gamma = 2.5$. See Table 2 for a comparison of flux and luminosity conversions using power-laws versus more complex spectral models. 
}
\end{flushleft}
\end{table*}  



\begin{table*}
\caption{Best-fitting parameters of the EPIC spectra from 2012 and 2017, and the ACIS spectrum from 2016} 
\vspace{-0.5cm}
\label{sev}
\begin{center}  
\tiny
\begin{tabular}{lccc} 
 \hline 
\hline \\[-3pt]
  Model Parameters      &       \multicolumn{3}{c}{Values} \\
  & 2012        &       2016        &        2017  \\[5pt]
\hline \\[-3pt]
\multicolumn{4}{c}{{\it tbabs} $\times$ {\it tbabs} $\times$ ({\it po} $+$ {\it gaussian})}     \\ [3pt]
\hline \\[-3pt]
$N_{\rm {H,Gal}}$   ($10^{22}$ cm$^{-2}$)    &    [0.20]      &    [0.20]     &       [0.20]   \\[4pt]
$N_{\rm {H,int}}$   ($10^{22}$ cm$^{-2}$)    &  $0.14^{+0.03}_{-0.03}$        &     $0.20^{+0.13}_{-0.12}$     &   $0.23 ^{+0.07}_{-0.06}$     \\[4pt]
$\Gamma$ (keV)    &   $2.17^{+0.07}_{-0.06}$       &    $2.63^{+0.22}_{-0.20}$    & $3.50^{+0.31}_{-0.26}$ \\[4pt]
$N_{\rm {po}}$  ($10^{-5}$ ph keV$^{-1}$ cm$^{-2}$ s$^{-1}$ at 1 keV)     &   $5.8^{+0.4}_{-0.4}$      &   $14.2^{+3.9}_{-2.9}$     &  $ 6.3^{+1.4}_{-1.1}$  \\[4pt]
$E_{\rm {line}}$  (keV)        &  $0.61^{+0.02}_{-0.02}$    &  --    &    $0.66^{+0.01}_{-0.01}$    \\[4pt]
$\sigma_{\rm {line}}$  (keV) & [0] & -- & $<0.030$ \\[4pt]
$N_{\rm {line}}$ ($10^{-5}$ ph cm$^{-2}$ s$^{-1}$) &   $1.2^{+1.2}_{-0.7}$       &      --          &  $ 2.7^{+2.3}_{-1.4}$\\[4pt]
$\chi^2/$dof      &    $346.8/230$ (1.51)      &    $68.0/61$  (1.12)      &        $89.3/64$ (1.40)  \\[4pt]

$f_{0.3-10}$ ($10 ^{-13}$ erg cm$^{-2}$ s$^{-1}$)$^a$   &  $1.68^{+0.06}_{-0.06}$   &  $2.49^{+0.22}_{-0.20}$    &  $  0.62^{+0.04}_{-0.04}$   \\[4pt]

$L_{0.3-10}$ ($10 ^{39}$ erg cm$^{-2}$ s$^{-1}$)$^b$  & $2.23^{+0.16}_{-0.14}$ &  $4.86^{+1.50}_{-0.97}$ & $ 3.10^{+1.45}_{-0.84} $   \\[4pt]

\hline \\[-3pt]
\multicolumn{4}{c}{{\it tbabs} $\times$ {\it tbabs} $\times$ ({\it bbodyrad} $+$ {\it diskbb} $+$ {\it gaussian})}     \\ [3pt]
\hline \\[-3pt]
$N_{\rm {H,Gal}}$   ($10^{22}$ cm$^{-2}$)    &    [0.20]      &    [0.20]     &       [0.20]   \\[4pt]
$N_{\rm {H,int}}$   ($10^{22}$ cm$^{-2}$)    &  $0.08^{+0.12}_{-0.08}$        &     $<0.15$     &   $0.26 ^{+0.23}_{-0.18}$     \\[4pt]
$kT_{\rm bb}$ (keV)    &   $0.15^{+0.03}_{-0.02}$       &    $0.27^{+0.08}_{-0.09}$    & $0.15^{+0.03}_{-0.02}$ \\[4pt]
$N_{\rm {bb}}$  (km$^2$)$^c$     &   $11.3^{+41.5}_{-9.2}$      &   $1.7^{+8.7}_{-1.0}$     &  $ 89^{+908}_{-79}$  \\[4pt]
$kT_{\rm in}$ (keV)        &    $1.25^{+0.07}_{-0.07}$            &    $1.14^{+0.40}_{-0.20}$        &       $0.72 ^{+0.14}_{-0.11}$\\[4pt]
$N_{\rm {dbb}} $  ($10 ^{-3}$ km$^2$)$^d$ & $3.7^{+1.1}_{-0.9}$     &   $6.3^{+10.0}_{-4.8}$        &     $13.3^{+19.7}_{-8.1} $   \\[4pt]
$E_{\rm {line}}$  (keV)        &  $0.61^{+0.04}_{-0.04}$    &  --    &    $0.66^{+0.02}_{-0.02}$    \\[4pt]
$\sigma_{\rm {line}}$  (keV) & [0] & -- & $<0.033$ \\[4pt]
$N_{\rm {line}}$ ($10^{-5}$ ph cm$^{-2}$ s$^{-1}$) &   $0.48^{+1.44}_{-0.26}$       &      --          &  $ 3.2^{+13.5}_{-2.1}$\\[4pt]
$\chi^2/$dof      &    $235.5/228$ (1.03)      &    $68.3/59$  (1.16)      &        $72.5/62$ (1.17)  \\[4pt]

$f_{0.3-10}$ ($10 ^{-13}$ erg cm$^{-2}$ s$^{-1}$)$^a$   &  $1.57^{+0.05}_{-0.05}$   &  $2.28^{+0.11}_{-0.10}$    &  $  0.61^{+0.04}_{-0.04}$   \\[4pt]

$L_{0.3-10}$ ($10 ^{39}$ erg cm$^{-2}$ s$^{-1}$)$^b$  & $1.75^{+0.76}_{-0.30}$ &  $2.22^{+0.76}_{-0.17}$ & $ 2.52^{+6.62}_{-1.44} $   \\[4pt]

\hline \\[-3pt]

\multicolumn{4}{c}{{\it tbabs} $\times$ {\it tbabs} $\times$ ({\it bbodyrad} $+$ {\it comptt} $+$ {\it gaussian})}      \\[3pt]
\hline\\[-3pt]
   $N_{\rm {H,Gal}}$   ($10^{22}$ cm$^{-2}$)    &    [0.20]      &    [0.20]     &       [0.20]   \\[4pt]
    $N_{\rm {H,int}}$   ($10^{22}$ cm$^{-2}$)        & $0.05^{+0.01}_{-0.01} $   & $0.23^{+0.33}_{-0.10} $  &  $0.28^{+0.03}_{-0.02} $ \\[4pt]

$kT_{\rm {bb}}$ (keV)           &  $0.16^{+0.01}_{-0.01} $  & $0.07^{+0.06}_{-0.07}$    &  $0.13^{+0.06}_{-0.01} $ \\[4pt]

$N_{\rm {bb}}$  (km$^2$)$^c$   & $7.3^{+13.7}_{-1.0}  $        &  (unconstrained) &    $117^{+258}_{-12} $   \\[4pt]

$kT_0$ (keV) $=$  $kT_{\rm {bb}}$     & [$0.16^{+0.01}_{-0.01}$]  &  [$0.07^{+0.06}_{-0.07}$]    &   [$0.13^{+0.06}_{-0.01} $]   \\[4pt]

$kT_e$ (keV)        &  $1.01^{+0.15}_{-0.10} $  &  $>0.98$   &       $0.65^{+0.89}_{\ast} $  \\[4pt]

$\tau$ & $13.7^{+0.4}_{-0.3} $  &  $5.0^{+3.4}_{-5.0} $ &  $13.6^{+7.3}_{-0.8} $ \\[4pt]
$N_{\rm{c}}$  ($10^{-5}$) &  $9.7^{+0.2}_{-0.2} $  & $63^{+606}_{\ast} $  &  $11.9^{+2.4}_{-1.1} $ \\[4pt]
$E_{\rm {line}}$  (keV)      &    $0.61^{+0.03}_{-0.04} $   &  -- &   $0.66^{+0.01}_{-0.02} $    \\[4pt]
$\sigma_{\rm {line}}$  (keV) & [0] & -- & $<0.025$ \\[4pt]
 $N_{\rm {line}}$ ($10^{-5}$ ph cm$^{-2}$ s$^{-1}$) &  $0.36^{+0.27}_{-0.27} $  & --  &   $3.7^{+1.4}_{-1.4} $   \\[4pt]
$\chi^2/$dof       & $234.6/227$ (1.03) & $64.1/58$ (1.11) &  $72.6/61$ (1.19)\\[4pt]
$f_{0.3-10}$ ($10 ^{-13}$ erg cm$^{-2}$ s$^{-1}$)$^a$  & $1.56^{+0.05}_{-0.05} $   & $2.59^{+1.85}_{-0.32} $ & $0.61^{+0.04}_{-0.04} $ \\[4pt]
$L_{0.3-10}$ ($10 ^{39}$ erg cm$^{-2}$ s$^{-1}$)$^b$  & $1.63^{+0.05}_{-0.04} $  & ($\gtrsim3.8$)$^e$ & $2.83^{+2.03}_{-0.08} $   \\[4pt]



\hline \\[-3pt]
\multicolumn{4}{c}{{\it tbabs} $\times$ {\it tbabs} $\times$ ({\it diskpbb} $+$ {\it gaussian})}      \\[3pt]
\hline\\[-3pt]
   $N_{\rm {H,Gal}}$   ($10^{22}$ cm$^{-2}$)    &    [0.20]      &    [0.20]     &       [0.20]   \\[4pt]
   $N_{\rm {H,int}}$   ($10^{22}$ cm$^{-2}$)   &  $<0.02$              &  $ 0.03^{+0.06}_{-0.03}$ &     $ 0.03^{+0.18}_{-0.02}$ \\[4pt]
   $kT_{\rm in}$ (keV)     &  $ 1.49^{+0.13}_{-0.11}$       &   $ 1.40^{+0.29}_{-0.31}$   &    $0.68 ^{+0.11}_{-0.09}$\\[4pt] 
   $p$    &   $ 0.60^{+0.01}_{-0.02}$            & $ 0.50^{+0.04}_{\ast}$     & $ 0.50^{+0.04}_{\ast}$\\[4pt]
   $N_{\rm {dpbb}} $  ($10^{-3}$ km$^2$)$^d$ &  $ 0.92^{+0.44}_{-0.31}$   &   $ 0.89^{+2.17}_{-0.53}$    & $6.7 ^{+8.1}_{-3.4}$  \\[4pt]
   $E_{\rm {line}}$  (keV)    &  $ 0.61^{+0.04}_{-0.04}$   & --  &    $ 0.67^{+0.01}_{-0.02}$   \\[4pt] 
$\sigma_{\rm {line}}$  (keV) & [0] & -- & $<0.029$ \\[4pt]
   $N_{\rm {line}}$ ($10^{-5}$ ph cm$^{-2}$ s$^{-1}$) &  $0.28^{+0.23}_{-0.19}$  & --  &  $ 0.80^{+0.45}_{-0.36}$  \\[4pt]
   $\chi^2/$dof     &      $252.7 /229$ (1.10)          &       $ 65.9/60$ (1.10)        &       $102.5/63$ (1.63) \\[4pt]
   $f_{0.3-10}$ ($10 ^{-13}$ erg cm$^{-2}$ s$^{-1}$)$^a$ & $ 1.58^{+0.06}_{-0.05}$ & $ 2.41^{+0.19}_{-0.20}$  &$ 0.59^{+0.04}_{-0.04}$ \\[4pt]
   $L_{0.3-10}$ ($10 ^{39}$ erg cm$^{-2}$ s$^{-1}$)$^b$ & $ 1.50^{+0.06}_{-0.05}$ & $ 1.71^{+0.13}_{-0.14}$ & $ 0.93^{+0.06}_{-0.06}$  \\[4pt]

\hline 
\vspace{-0.55cm}
\end{tabular}
\end{center}
\begin{flushleft} 
\tiny{
$^a$: observed fluxes in the 0.3--10 keV band\\
$^b$: for all spectral models, the de-absorbed luminosities $L_{0.3-10}$ (0.3--10 keV band) were defined as $4\pi d^2$ times the de-absorbed fluxes \\
$^c$: $N_{\rm {bb}} = (R_{\rm{bb}}/D_{10})^2$ where $R_{\rm{bb}}$ is the source radius in km and $D_{10}$ is the distance to the source in units of 10 kpc (here, $D_{10} = 770$).\\
$^d$: $N_{\rm {dbb}} = (R_{\rm{in}}/D_{10})^2 \cos \theta$, where $R_{in}$ is the apparent inner disk radius in km, $D_{10}$ the distance to the source in units of 10 kpc, and $\theta$ is our viewing angle ($\theta = 0$ is face-on). $N_{\rm {dpbb}}$ is defined exactly as $N_{\rm {dbb}}$.\\
$^e$: upper limit unconstrained because of the degeneracy between absorption column density and seed blackbody luminosity}
\end{flushleft}
\end{table*}


\begin{table*}
\caption{Alternative set of models for the 2017 EPIC spectra}
\vspace{-0.3cm}
\label{sev}
\begin{center}  
\tiny
\begin{tabular}{lc} 
 \hline 
\hline \\[-3pt]
  Model Parameters      &       Values  \\[5pt]
\hline \\[-3pt]
\multicolumn{2}{c}{{\it tbabs} $\times$ {\it tbabs} $\times$ ({\it bbodyrad} $+$ {\it diskbb} $+$ {\it vapec})}     \\ [3pt]
\hline \\[-3pt]
$N_{\rm {H,Gal}}$   ($10^{22}$ cm$^{-2}$)    &      [0.20]   \\[4pt]
$N_{\rm {H,int}}$   ($10^{22}$ cm$^{-2}$)    &   $0.13^{+0.19}_{-0.13}$     \\[4pt]
$kT_{\rm bb}$ (keV)     & $0.14^{+0.06}_{-0.04}$ \\[4pt]
$N_{\rm {bb}}$  (km$^2$)$^a$     &   $ 22^{+358}_{-21}$  \\[4pt]
$kT_{\rm in}$ (keV)        &  $0.72 ^{+0.14}_{-0.12}$\\[4pt]
$N_{\rm {dbb}} $  ($10 ^{-3}$ km$^2$)$^b$ &     $7.5^{+9.5}_{-4.3} $   \\[4pt]
$kT_{\rm vapec}$ (keV)    & $0.41^{+0.09}_{-0.10}$ \\[4pt]
O abundance (solar units)    &   $>$10  \\[4pt]
Ne = Mg = Si abundances (solar units)    &   $>$10  \\[4pt]
Other abundances (solar units)    &     [1]  \\[4pt]
$N_{\rm {vapec}}$ ($10^{-6}$)$^c$    &  $ 0.07^{+4.3}_{-0.03}$\\[4pt]
$\chi^2/$dof    &        $68.6/61$ (1.13)  \\[4pt]

$f_{0.3-10}$ ($10 ^{-13}$ erg cm$^{-2}$ s$^{-1}$)$^d$    &  $  0.62^{+0.04}_{-0.04}$   \\[4pt]

$L_{0.3-10}$ ($10 ^{39}$ erg cm$^{-2}$ s$^{-1}$)$^e$   & $ 1.32^{+2.51}_{-0.57} $   \\[4pt]

\hline \\[-3pt]

\multicolumn{2}{c}{{\it tbabs} $\times$ {\it tbabs} $\times$ ({\it bbodyrad} $+$ {\it comptt} $+$ {\it vapec})}      \\[3pt]
\hline\\[-3pt]
   $N_{\rm {H,Gal}}$   ($10^{22}$ cm$^{-2}$)    &       [0.20]   \\[4pt]
    $N_{\rm {H,int}}$   ($10^{22}$ cm$^{-2}$)   &  $0.12^{+0.16}_{-0.02} $ \\[4pt]

$kT_{\rm {bb}}$ (keV)          &  $0.15^{+0.01}_{-0.02} $ \\[4pt]

$N_{\rm {bb}}$  (km$^2$)$^c$   &     $17.5^{+39.9}_{-2.8} $   \\[4pt]

$kT_0$ (keV) $=$  $kT_{\rm {bb}}$     &   [$0.15^{+0.01}_{-0.02} $]   \\[4pt]

$kT_e$ (keV)     &       $0.59^{+0.46}_{\ast} $  \\[4pt]

$\tau$  &  $26.4^{+4.1}_{-3.1} $ \\[4pt]
$N_{\rm{c}}$  ($10^{-5}$)  &  $4.3^{+0.7}_{-0.5} $ \\[4pt]
$kT_{\rm vapec}$ (keV)    & $0.43^{+0.05}_{-0.04}$ \\[4pt]
O abundance (solar units)    &   $385^{+155}_{-140}$  \\[4pt]
Ne = Mg = Si abundances (solar units)    &   $230^{+110}_{-95}$  \\[4pt]
Other abundances (solar units)    &     [1]  \\[4pt]
 $N_{\rm {vapec}}$ ($10^{-6}$)$^c$   &   $0.18^{+0.65}_{-0.05} $   \\[4pt]
$\chi^2/$dof &  $68.4/60$ (1.14)\\[4pt]

$f_{0.3-10}$ ($10 ^{-13}$ erg cm$^{-2}$ s$^{-1}$)$^d$ & $0.61^{+0.04}_{-0.04} $ \\[4pt]

$L_{0.3-10}$ ($10 ^{39}$ erg cm$^{-2}$ s$^{-1}$)$^e$   & $1.23^{+2.71}_{-0.07} $   \\[4pt]



\hline \\[-3pt]
\multicolumn{2}{c}{{\it tbabs} $\times$ {\it tbabs} $\times$ ({\it diskpbb} $+$ {\it vapec})}      \\[3pt]
\hline\\[-3pt]
   $N_{\rm {H,Gal}}$   ($10^{22}$ cm$^{-2}$)    &       [0.20]   \\[4pt]
   $N_{\rm {H,int}}$   ($10^{22}$ cm$^{-2}$) &  $0.05^{+0.06}_{-0.05}$ \\[4pt]
   $kT_{\rm in}$ (keV)     &    $0.93 ^{+0.23}_{-0.16}$\\[4pt] 
   $p$       & $ 0.50^{+0.12}_{\ast}$\\[4pt]
   $N_{\rm {dpbb}} $  ($10^{-3}$ km$^2$)$^b$   & $1.2^{+6.7}_{-0.6}$  \\[4pt]
$kT_{\rm vapec}$ (keV)    & $0.40^{+0.07}_{-0.04}$ \\[4pt]
O abundance (solar units)    &   $>$5.5  \\[4pt]
Ne = Mg = Si abundances (solar units)    &   $>$6.3  \\[4pt]
Other abundances (solar units)    &     [1]  \\[4pt]
   $N_{\rm {vapec}}$ ($10^{-6}$)$^c$  &  $2.8^{+4.5}_{-2.7}$  \\[4pt]
   $\chi^2/$dof      &       $71.7/62$ (1.16) \\[4pt]
   
   $f_{0.3-10}$ ($10^{-13}$ erg cm$^{-2}$ s$^{-1}$)$^d$   &$ 0.62^{+0.04}_{-0.04}$ \\[4pt]
   
   $L_{0.3-10}$ ($10 ^{39}$ erg cm$^{-2}$ s$^{-1}$)$^e$  & $ 0.95^{+0.23}_{-0.20}$  \\[4pt]

\hline 
\end{tabular}
\end{center}
\begin{flushleft} 
\tiny{
$^a$: $N_{\rm {bb}} = (R_{\rm{bb}}/D_{10})^2$ where $R_{\rm{bb}}$ is the source radius in km and $D_{10}$ is the distance to the source in units of 10 kpc (here, $D_{10} = 770$).\\
$^b$: $N_{\rm {dbb}} = (R_{\rm{in}}/D_{10})^2 \cos \theta$, where $R_{in}$ is the apparent inner disk radius in km, $D_{10}$ the distance to the source in units of 10 kpc, and $\theta$ is our viewing angle ($\theta = 0$ is face-on). $N_{\rm {dpbb}}$ is defined exactly as $N_{\rm {dbb}}$.\\
$^c$: $N_{\rm {vapec}} = {10^{-14}\over{4\pi\,d^2}}\int n_en_{\rm H}dV$, where $d$ is the angular diameter distance to the source (cm), and $n_e$ and $n_{\rm H}$ are the electron and hydrogen densities (cm$^{-3}$)\\
$^d$: observed fluxes in the 0.3--10 keV band\\
$^e$: for all spectral models, the de-absorbed luminosities $L_{0.3-10}$ (0.3--10 keV band) were defined as $4\pi d^2$ times the de-absorbed fluxes}
\end{flushleft}
\end{table*}

\begin{table*} 
\caption{{\it HST} observations of a candidate counterpart of CXOU J203451.1$+$601043}
 \label{five} 
\begin{center}  
\begin{tabular}{lccccc}  
\hline \hline \\[-3pt]   
 Observation Date   &   Detector & Filter             & Exposure Time    &      Apparent Brightness & Absolute Magnitude       \\
 &   &   & (s) & (Vegamag) & (Vegamag) \\[5pt]
\hline  \\[-3pt]
  2004-07-29 &  ACS-WFC   &   F814W     &  120    & --   & -- \\[3pt]
  \hline\\[-3pt]
  2016-10-26 &  ACS-WFC   &   F606W     &  2430    &  $26.35 \pm 0.15$   &  $-3.95 \pm 0.15$ \\[3pt]
  2016-10-26 &  ACS-WFC  & F814W     &   2570   &   $26.05 \pm 0.15$ &   $-3.90 \pm 0.15$ \\[3pt] 
 \hline\\[-3pt]
 2018-01-05 &  WFC3-UVIS & F555W      &  710      & -- &        --  \\[3pt]
 2018-01-05 &  WFC3-UVIS & F814W      &  780      &  -- &        --  \\[3pt]
        
\hline  

\end{tabular} 
\end{center}

\end{table*}  

\clearpage
\begin{figure*}[ht]
\includegraphics[width=16cm,clip]{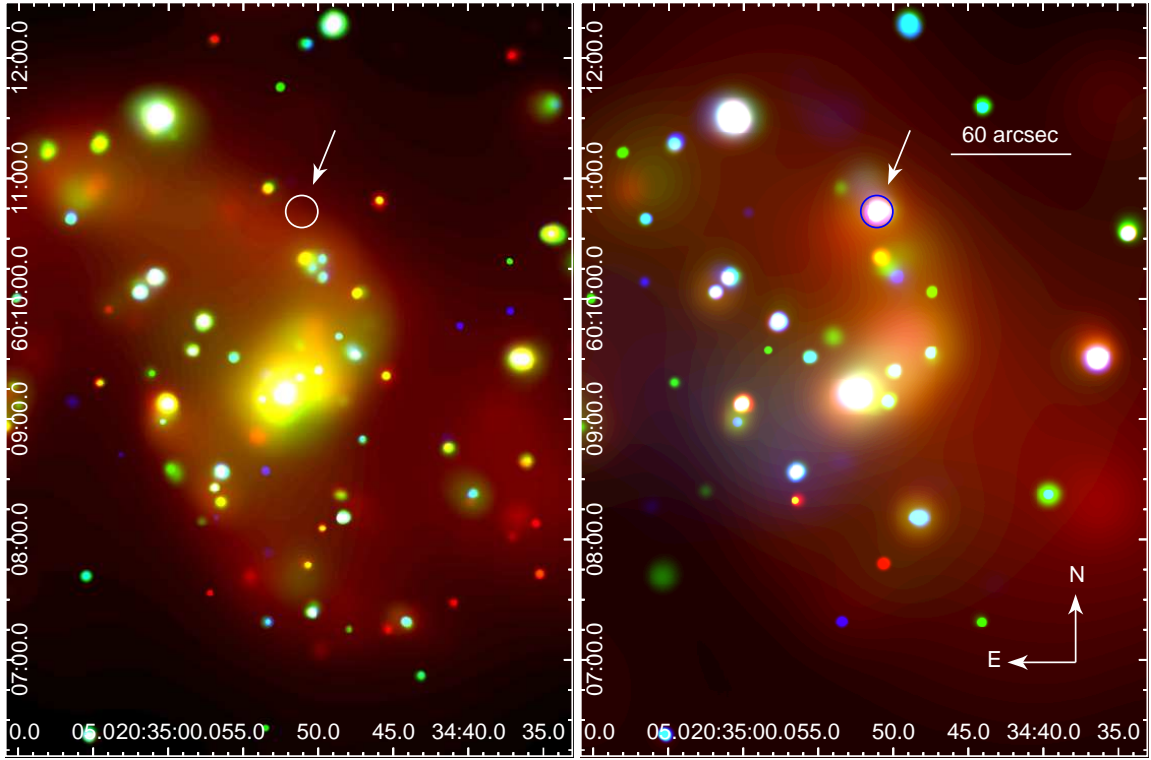}
\caption{Left panel: adaptively smoothed {\it Chandra}/ACIS image of NGC\,6946, based on the stacked data from 2001 to 2004. Red represents the 0.3--1 keV band, green is for 1--2 keV, and blue for 2--8 keV. Right panel: adaptively smoothed {\it Chandra}/ACIS image, based on the stacked data from 2012 to 2017, showing the appearance of the transient ULX investigated in this paper.}
\label{fig1}
\end{figure*}

\begin{figure*}[ht]
\includegraphics[width=16cm,clip]{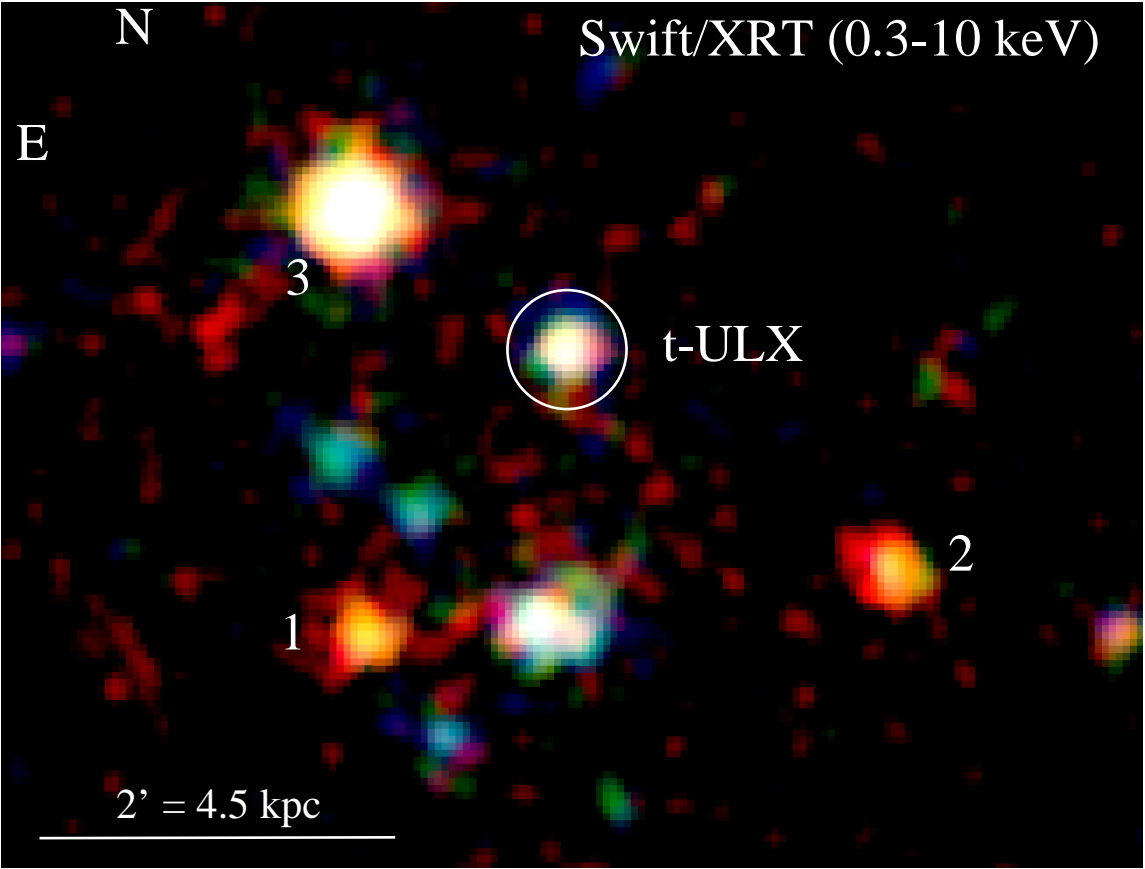}
\caption{Gaussian-smoothed {\it Swift}/XRT image of the field around the transient ULX CXOU J203451.1$+$601043 (labelled as t-ULX) in NGC\,6946, based on the stacked data from 2018 April to 2019 April, showing that the source is currently still ultraluminous. Red represents the 0.3--1 keV band, green is for 1--2 keV, and blue for 2--10 keV. The other bright off-nuclear sources labelled ``1'', ``2'' and ``3'' correspond to ULX-1, ULX-2 and ULX-3 in \cite{earnshaw19b} (see their Fig.~1); in particular, ULX-3 is the well-studied ULX inside the MF\,16 nebula.}
\label{fig2}
\end{figure*}

\begin{figure*}[ht]
\includegraphics[width=16cm,clip]{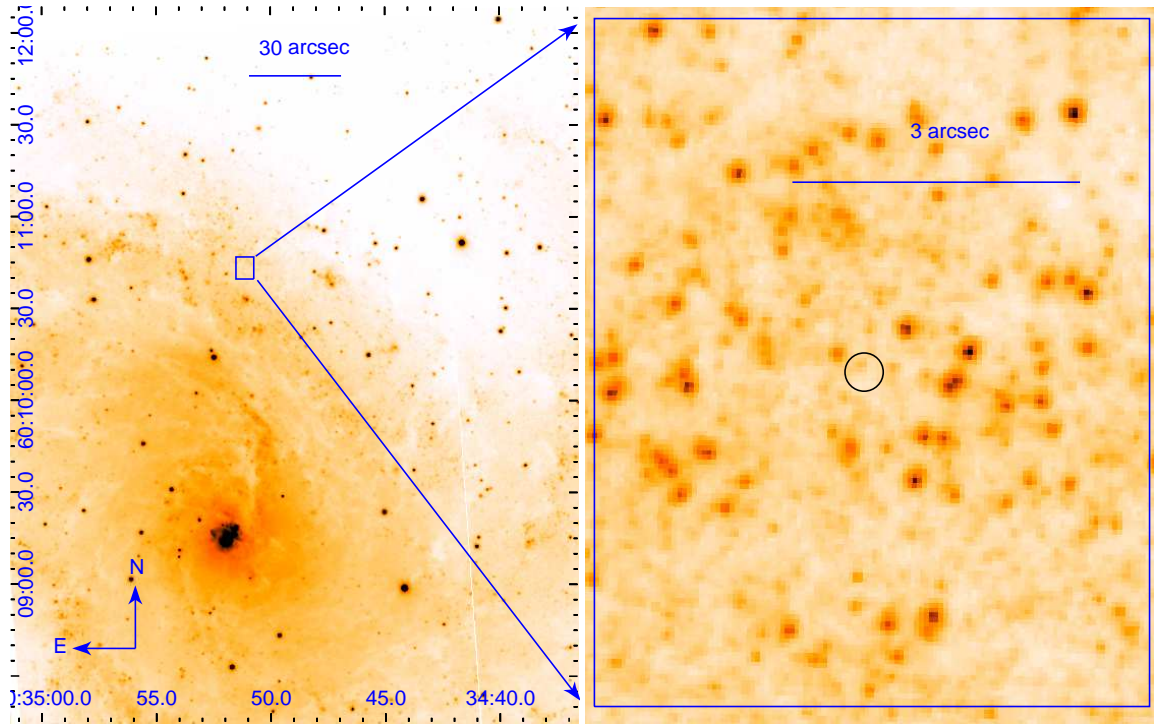}
\caption{Left panel: archival Gemini-North $i$-band image; the box marks the location of the transient ULX and is zoomed in on the right. Right panel: stellar field around CXOU J203451.1$+$601043, from an {\it HST}/ACS image in the F814W band. The blue circle shows the location of the transient ULX and has a 90\% error radius of 0$^{\prime\prime}$.2.}
\label{optical}
\end{figure*}


\begin{figure*}[ht]
\centering
\includegraphics[angle=270,width=15cm]{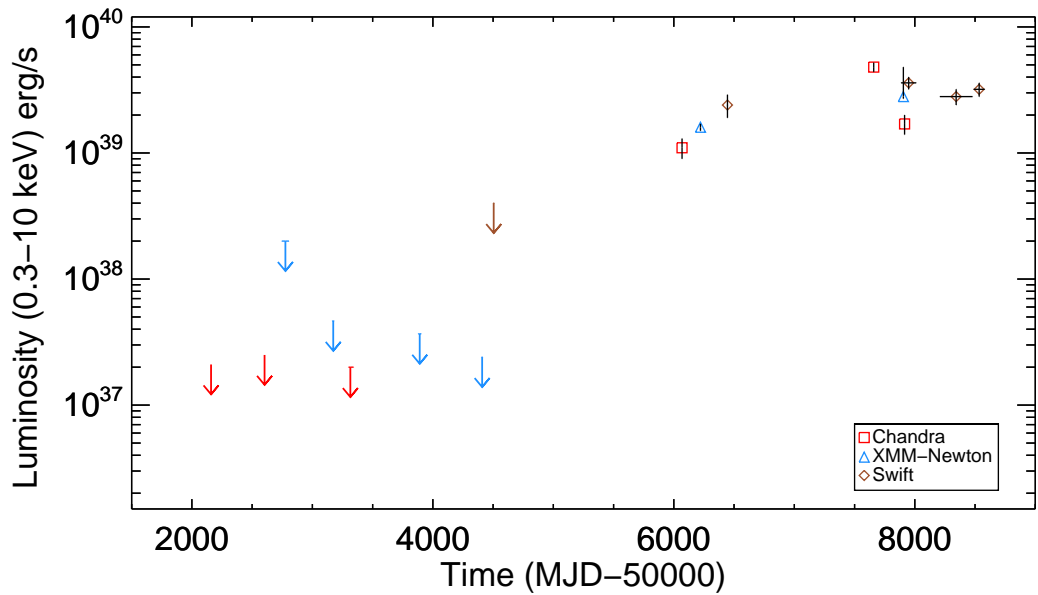}
\vspace{0.3cm}
\caption{Long-term X-ray luminosity evolution of CXOU J203451.1$+$601043 in the 0.3--10 keV band (data from Table 1). When sufficient counts were available, we estimated the luminosities from detailed spectral modelling of individual observations (Table 2); when only few counts were available, or for non-detection limits, we used a fiducial power-law model with photon index $\Gamma = 2.5$ and intrinsic column density $N_{\rm H} = 2 \times 10^{21}$ cm$^{-2}$, which approximates the average spectrum of the source. Each of the {\it Swift} data points in this plot is a stack of several short observations over intervals of a few months.}
\label{long-lc}
\end{figure*}


\begin{figure*}[ht]
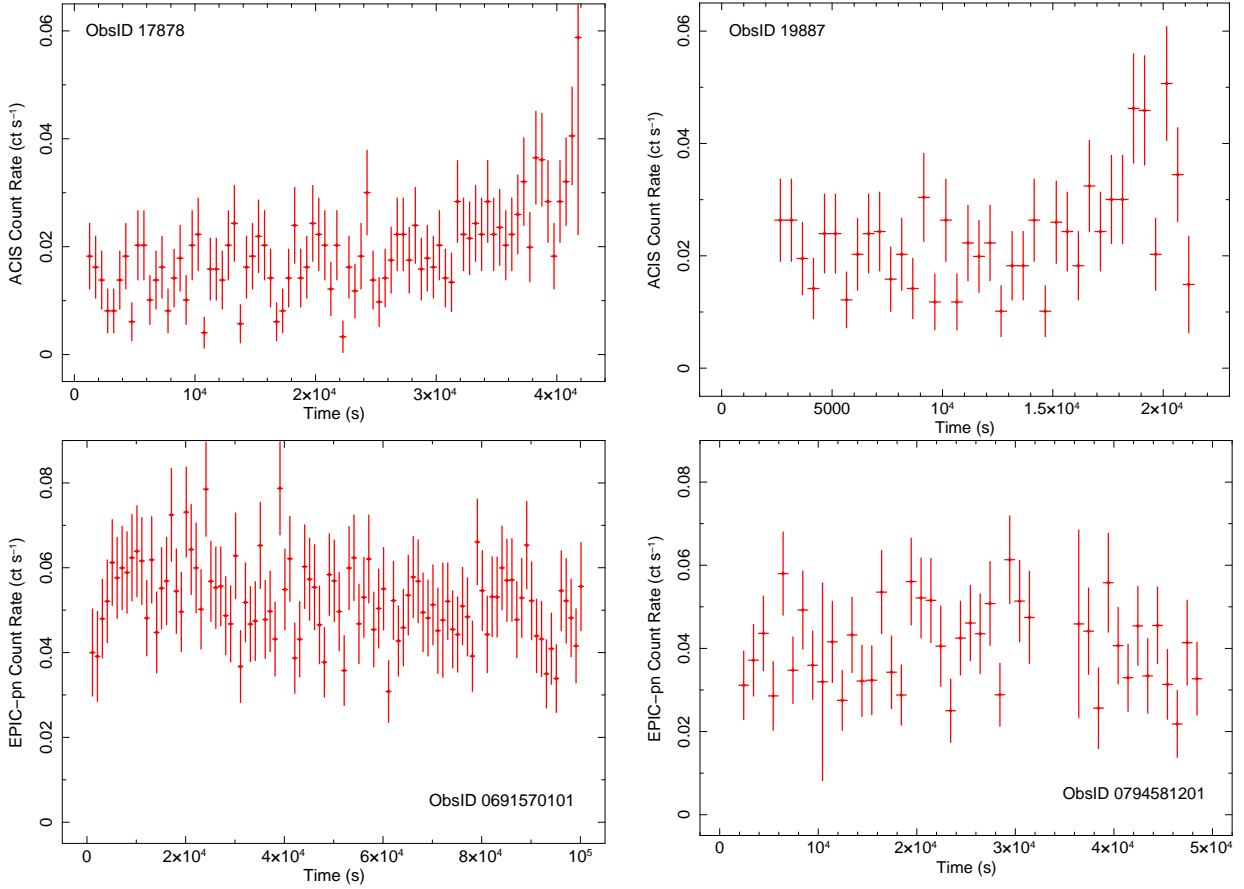

\includegraphics[angle=-90,width=8.4cm]{f5a.eps}
\includegraphics[angle=-90,width=8.2cm]{f5b.eps}\\
\includegraphics[angle=-90,width=8.4cm]{f5c.eps}
\includegraphics[angle=-90,width=8.4cm]{f5d.eps}

\caption{Top left panel: background-subtracted {\it Chandra}/ACIS-S light curve from the first of the two exposures on 2016 September 28, binned to 500 s; it shows moderate intra-observational variability. Top right panel: as in the top left panel, for the second ACIS-S exposure on 2016 September 28. Bottom left panel: background-subtracted {\it XMM-Newton}/EPIC-pn light curve from 2012 October 21, binned to 1000 s. Bottom right panel: as in the bottom left panel, for the EPIC-pn observation of 2017 June 1.}
\label{short-lc}
\end{figure*}

\begin{figure*}[ht]
\centering
\includegraphics[angle=-90,width=12cm]{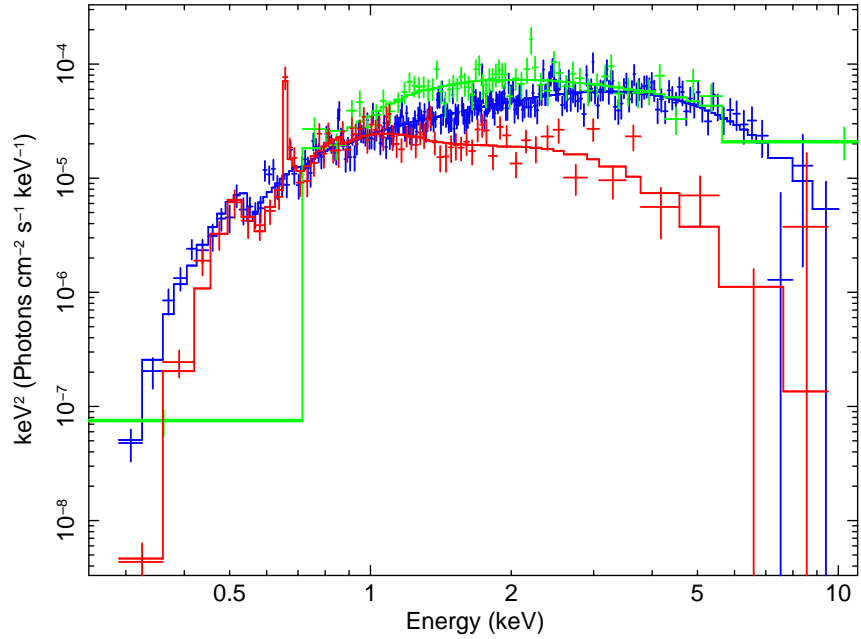}
\caption{Unfolded X-ray spectra of CXOU J203451.1$+$601043 at three different epochs, based on the best-fitting Comptonization models listed in Table 2 ({\it tbabs} $\times$ {\it tbabs} $\times$ ({\it bbodyrad} $+$ {\it comptt} $+$ {\it gaussian})). Blue datapoints are for the 2012 {\it XMM-Newton} spectrum; green datapoints for the 2016 {\it Chandra} spectrum; red datapoints for the 2017 {\it XMM-Newton} spectrum (including a strong oxygen emission line). }
\label{ufs}
\end{figure*}

\begin{figure*}[ht]
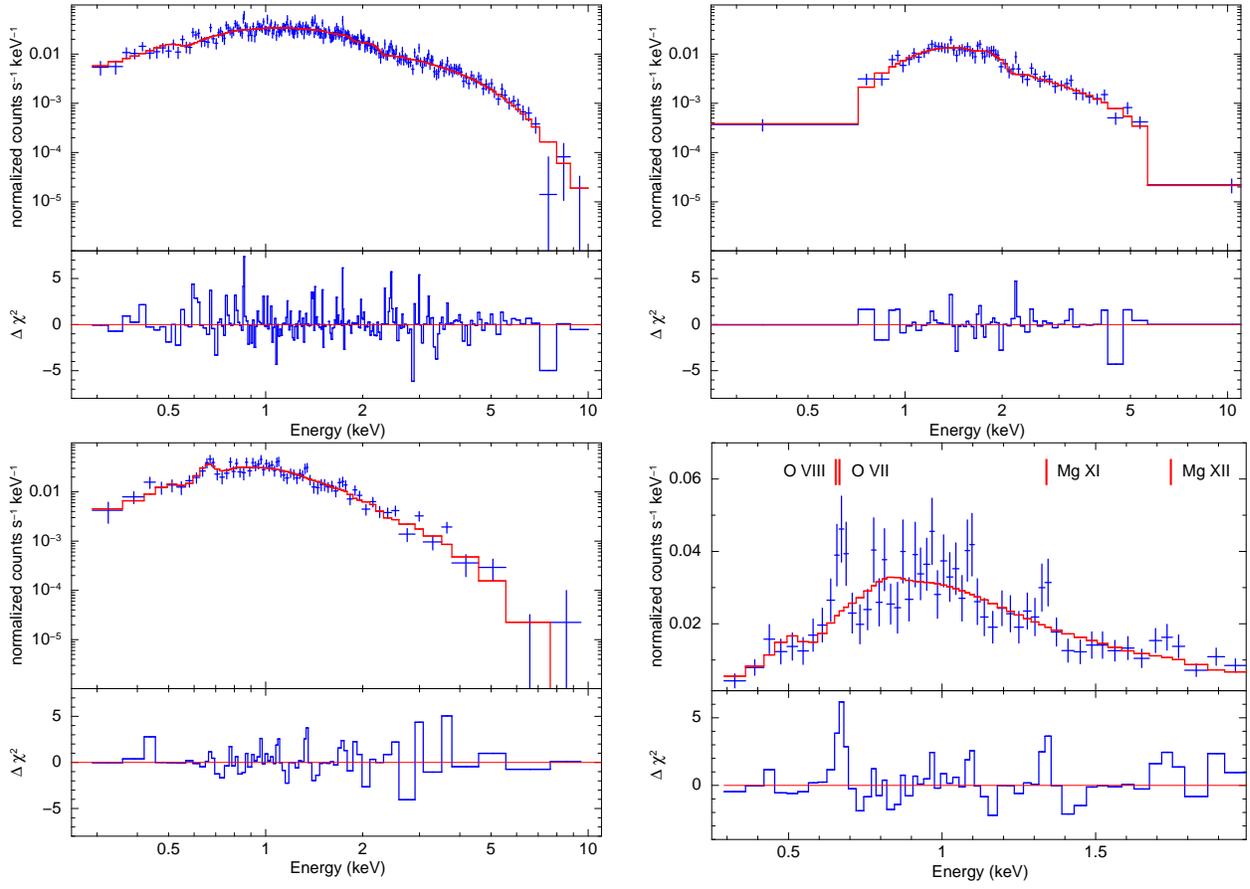

\centering
\includegraphics[angle=-90,width=8.4cm]{f7a.eps}
\includegraphics[angle=-90,width=8.4cm]{f7b.eps}
\includegraphics[angle=-90,width=8.4cm]{f7c.eps}
\includegraphics[angle=-90,width=8.4cm]{f7d.eps}
\caption{Top left panel: best-fitting spectrum and $\chi^2$ residuals for the 2012 {\it XMM-Newton}/EPIC dataset (pn and MOS combined), fitted with a Comptonization model (see Table 2 for the fit parameters). Top right panel: as in the top left panel, for the 2016 {\it Chandra}/ACIS-S spectrum. Bottom left panel: as in the top left panel, for the 2017 {\it XMM-Newton}/EPIC spectrum; notice the strong line at 0.66 keV, fitted with a Gaussian. Bottom right panel: zoomed-in view of the soft X-ray band for the 2017 {\it XMM-Newton}/EPIC spectrum, fitted this time only with a continuum model, without the addition of any lines; systematic residuals are clearly significant.}
\label{spec}
\end{figure*}

\begin{figure*}[ht]
\centering
\includegraphics[angle=0,width=12cm]{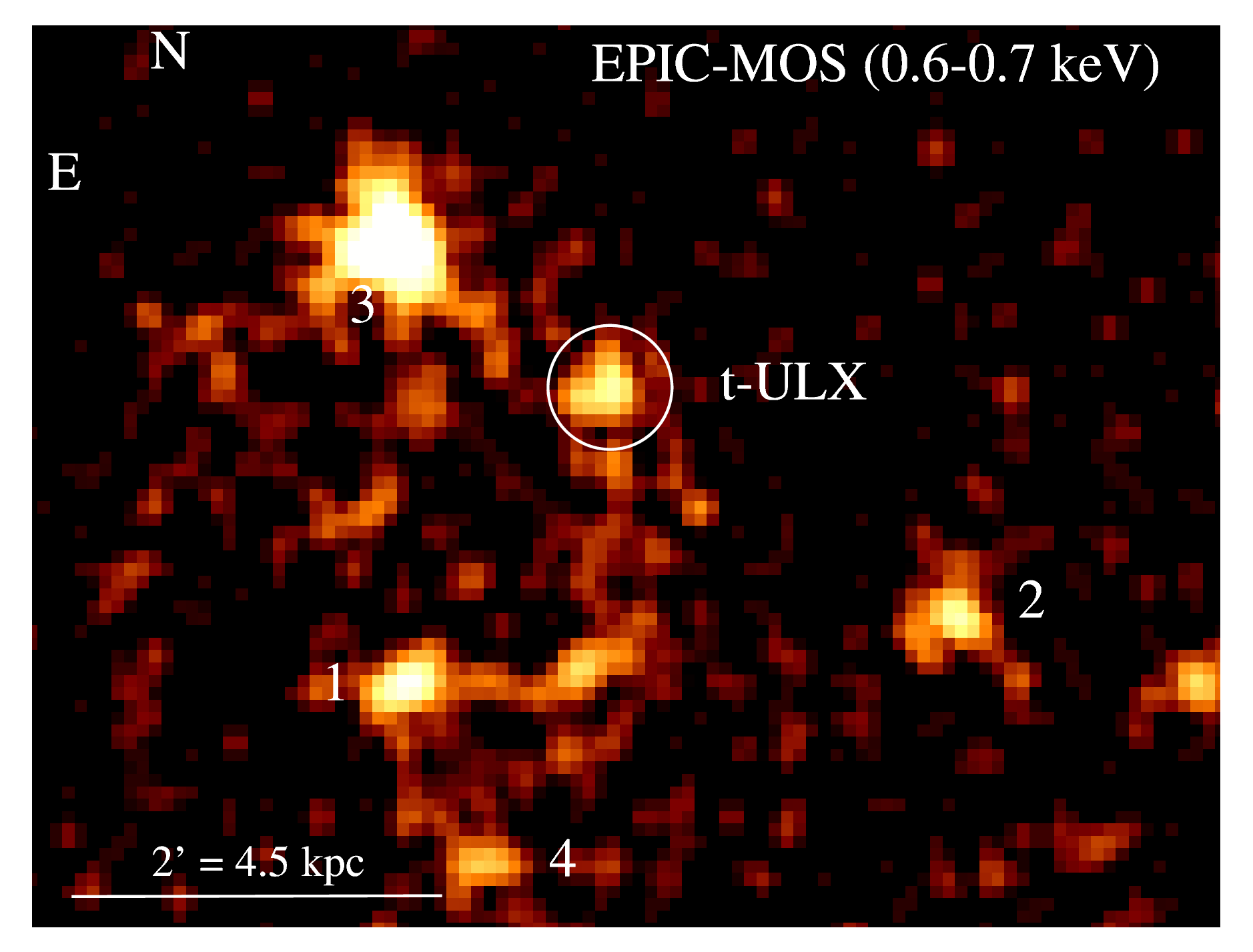}
\includegraphics[angle=0,width=12cm]{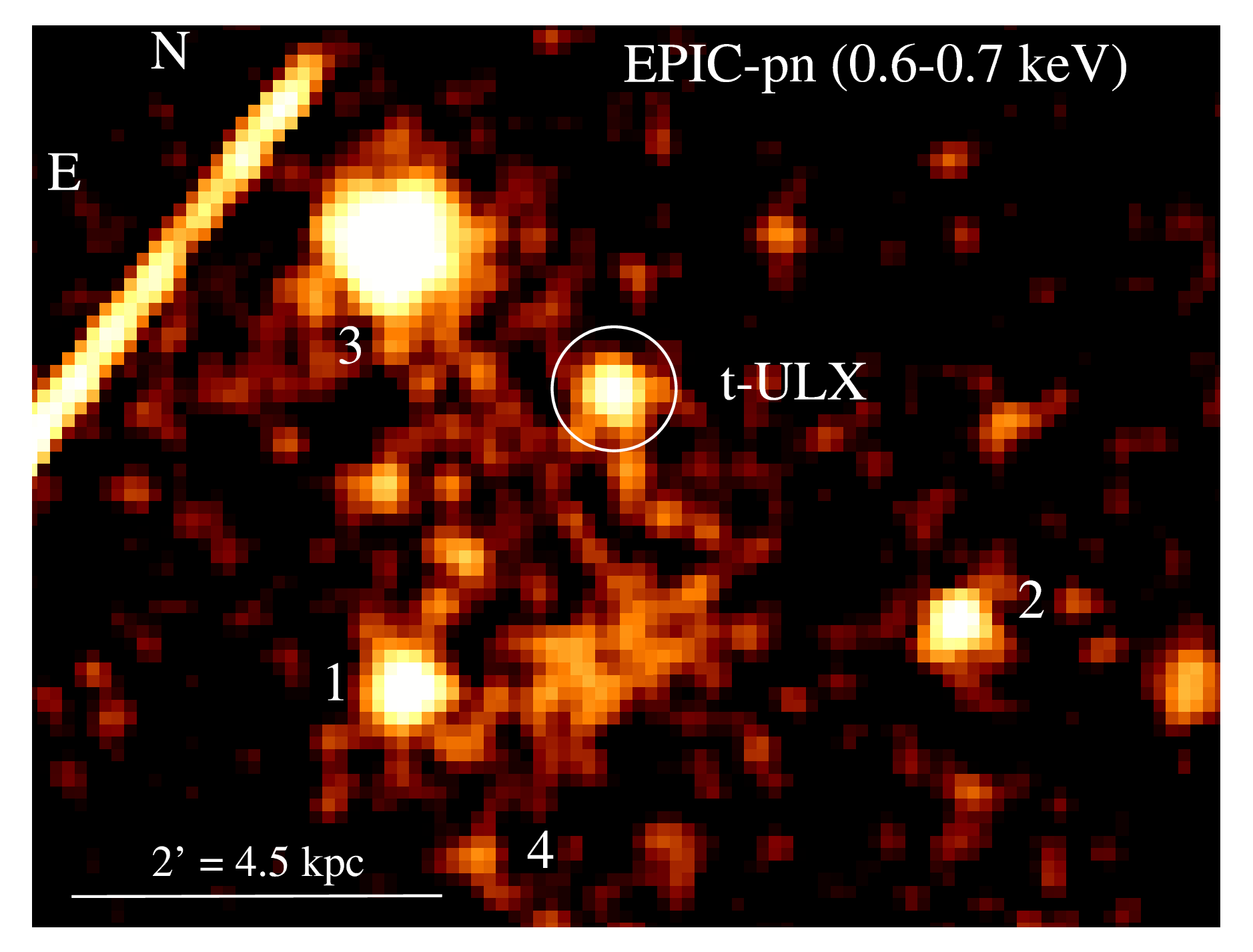}
\caption{Top panel: {\it XMM-Newton}/EPIC-MOS image from the 2017 dataset, filtered to the 0.60--0.70 keV band; it shows that the O {\footnotesize{VIII}} line emission is associated with the point-like ULX (labelled as t-ULX), and is not due to the contamination from diffuse hot gas. The other bright off-nuclear sources labelled ``1'', ``2'', ``3'' and ``4'' correspond to ULX-1, ULX-2, ULX-3 and ULX-4 in \cite{earnshaw19b}. 
Bottom panel: as in the top panel, for the EPIC-pn image, consistent with the MOS image; it shows that the line is not an instrumental artifact in one of the EPIC detectors. (The bright streak on the left is a bad column, flagged out for spectral analysis.)}
\label{oxygen}
\end{figure*}

\begin{figure*}[ht]
\centering
\includegraphics[angle=0,width=12cm]{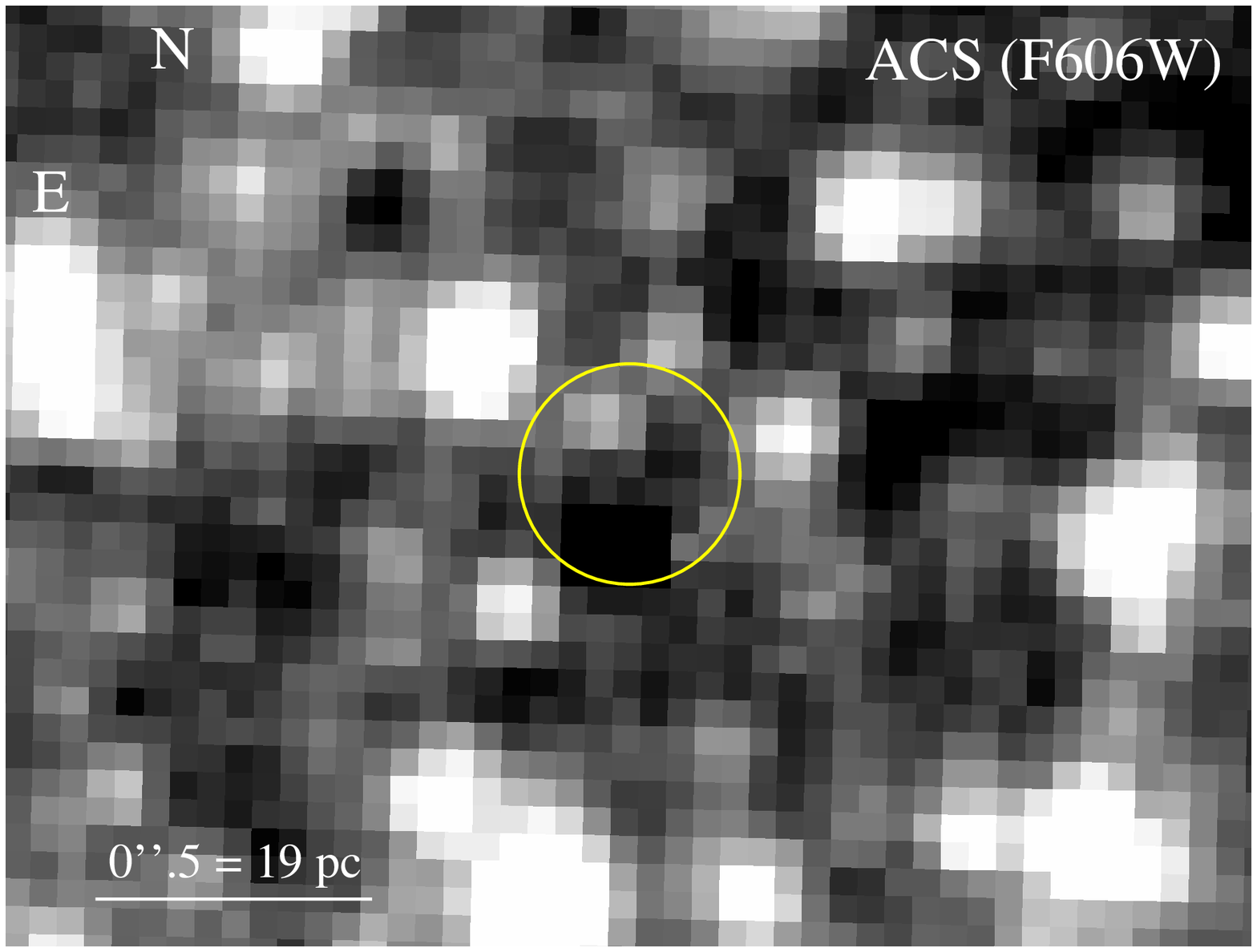}
\includegraphics[angle=0,width=12cm]{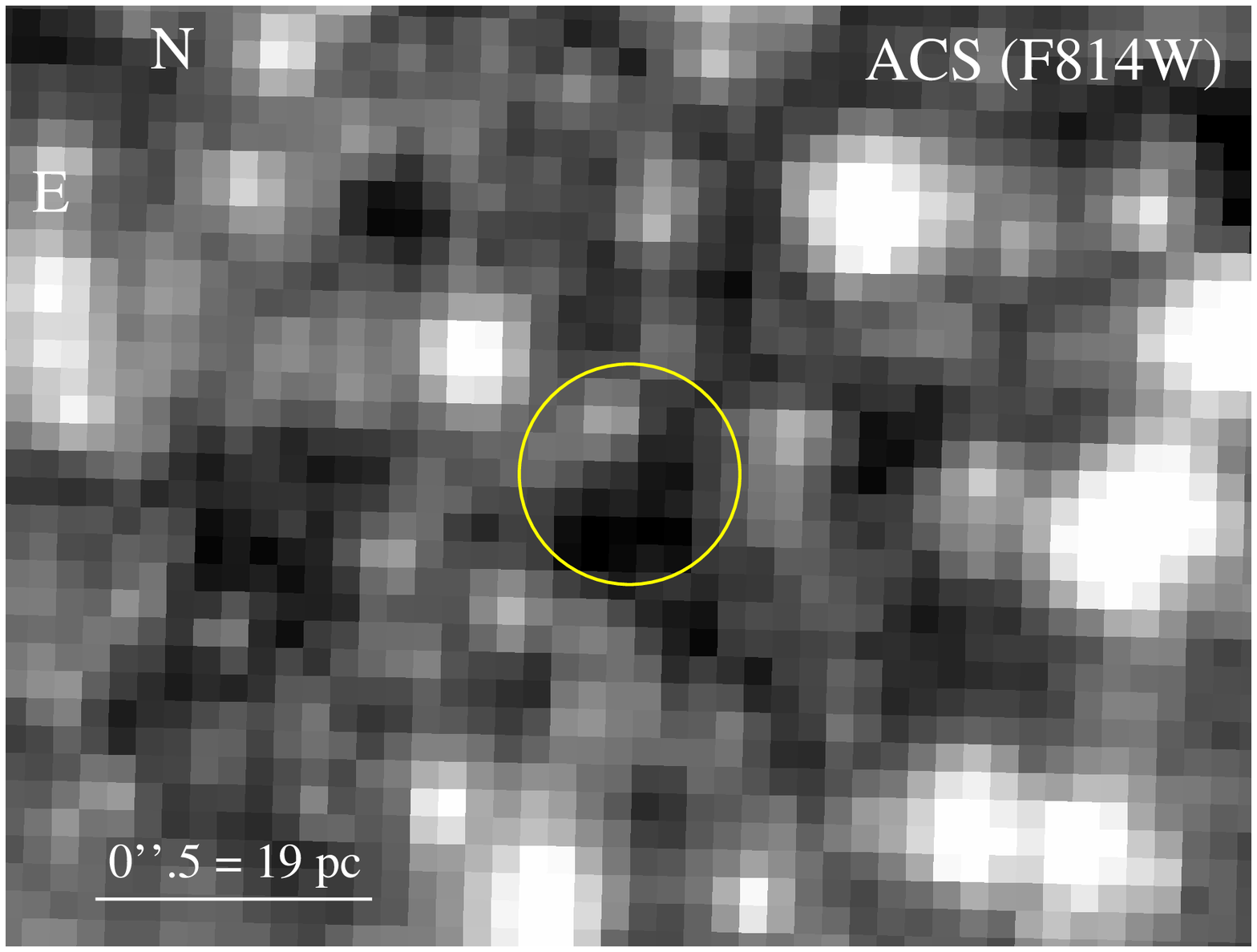}
\caption{Top panel: 2016 {\it HST}/ACS image in the F606W band. The yellow circle represents the 90\% confidence limit of 0$^{\prime\prime}$.2 for the ULX position. The only source marginally detected inside the circle has an apparent brightness of $\approx$26.4 mag and an absolute magnitude of $\approx -4.0$ mag.  Bottom panel: 2016 {\it HST}/WFC3 image in the F814W band.}
\label{counterparts}
\end{figure*}

\end{document}